\pdfoutput=1
\documentclass[11pt,twoside,a4paper,cmspaper,final,collab]{cms-tdr}
\def\svnVersion{a8e4dc6}\def\svnDate{2023/07/28}\def\cmsCernNoTag{CERN-EP-2022-112}\def\cmsCernDate{\today}\def\cmsMessage{Published in the Journal of High Energy Physics as \href{http://dx.doi.org/10.1007/JHEP07(2023)077}{\doi{10.1007/JHEP07(2023)077}.}}
\begin{document}\cmsNoteHeader{TOP-21-008}

\newlength\cmsTabSkip\setlength{\cmsTabSkip}{1.5ex}

\newcommand{\Wjets}{\ensuremath{\PW\text{+jets}}\xspace}
\newcommand{\Zjets}{\ensuremath{\PZ\text{+jets}}\xspace}
\newcommand{\mumu}{\ensuremath{\MM}\xspace}
\newcommand{\ee}{\ensuremath{\EE}\xspace}
\newcommand{\emu}{\ensuremath{\Pepm\PGmmp}\xspace}
\newcommand{\FXFX}{\ensuremath{\text{FxFx}}\xspace}

\newcommand{\PYTHIAEight}{\ensuremath{\PYTHIA8}\xspace}
\newcommand{\POWHEGPYTHIAEight}{\ensuremath{\POWHEG\text{+}\PYTHIAEight}\xspace}

\newcommand{\ttbarjet}{\ensuremath{\ttbar\text{+jet}}\xspace}
\newcommand{\ttbarnojet}{\ensuremath{\ttbar\text{+0~jet}}\xspace}
\newcommand{\mtop}{\ensuremath{m_{\PQt}}\xspace}
\newcommand{\mtopMC}{\ensuremath{\mtop^{\text{MC}}}\xspace}
\newcommand{\mtpole}{\ensuremath{\mtop^{\text{pole}}}\xspace}
\newcommand{\lumivalueSix}{36.3\fbinv}
\newcommand{\hdamp}{\ensuremath{h_{\text{damp}}}\xspace}
\newcommand{\muR}{\ensuremath{\mu_{\mathrm{R}}}\xspace}
\newcommand{\muF}{\ensuremath{\mu_{\mathrm{F}}}\xspace}
\newcommand{\chisq}{\ensuremath{\chi^2}\xspace}
\newcommand{\mll}{\ensuremath{m_{\Pell\Pell}}\xspace}
\newcommand{\mlb}{\ensuremath{m_{\Pell\PQb}}\xspace}
\newcommand{\dnnresponse}{\ensuremath{R_{\text{NN}}}\xspace}
\newcommand{\mlbmin}{\ensuremath{\mlb^{\text{min}}}\xspace}
\newcommand{\nj}{\ensuremath{N_{\text{jet}}}\xspace}
\newcommand{\nbj}{\ensuremath{N_{\PQb\text{\,jet}}}\xspace}
\newcommand{\Isorel}{\ensuremath{I_{\text{rel}}}\xspace}
\newcommand{\mzero}{\ensuremath{m_0}\xspace}
\newcommand{\pp}{\ensuremath{\Pp\Pp}\xspace}
\newcommand{\tW}{\ensuremath{\PQt\PW}\xspace}
\newcommand{\rhotrue}{\ensuremath{\rho_{\text{true}}}\xspace}
\newcommand{\rhoreco}{\ensuremath{\rho_{\text{reco}}}\xspace}
\newcommand{\Zmass}{\ensuremath{m_{\PZ}}\xspace}

\newcommand{\mttbarjet}{\ensuremath{m_{\ttbarjet}}\xspace}
\newcommand{\sigeta}{\ensuremath{\eta}\xspace}
\newcommand{\abseta}{\ensuremath{\abs{\eta}}\xspace}
\newcommand{\NN}{\ensuremath{\PGn\PAGn}\xspace}
\newcommand{\WW}{\ensuremath{\PWp\PWm}\xspace}

\newcommand{\sttj}{\ensuremath{s(\ttbarjet)}\xspace}
\newcommand{\sigttj}{\ensuremath{\sigma_{\ttbarjet}}\xspace}
\newcommand{\sigttjk}{\ensuremath{\sigttj^k}\xspace}
\newcommand{\sigttjkMC}{\ensuremath{\sigttj^k\text{(MC)}}\xspace}
\newcommand{\dsigttjdrho}{\ensuremath{\ddinline{\sigttj}{\rho}}\xspace}
\newcommand{\rhoklow}{\ensuremath{\rho_{\text{low}}^k}\xspace}
\newcommand{\rhokhigh}{\ensuremath{\rho_{\text{high}}^k}\xspace}

\newcommand{\likelihood}{\ensuremath{\mathcal{L}}\xspace}
\newcommand{\vi}{\ensuremath{v_i}\xspace}
\newcommand{\numi}{\ensuremath{n_i}\xspace}
\newcommand{\wj}{\ensuremath{w_j}\xspace}
\newcommand{\lm}{\ensuremath{\lambda_m}\xspace}
\newcommand{\sik}{\ensuremath{s_i^k}\xspace}
\newcommand{\bij}{\ensuremath{b_i^j}\xspace}
\newcommand{\vecl}{\ensuremath{\vec{\lambda}}\xspace}
\newcommand{\rk}{\ensuremath{r_k}\xspace}
\newcommand{\HTB}{\ensuremath{\HT^B}\xspace}

\newcommand{\hattheta}{\ensuremath{\hat{\theta}}\xspace}
\newcommand{\thetaz}{\ensuremath{\theta_0}\xspace}
\newcommand{\dtheta}{\ensuremath{\Delta\theta}\xspace}
\newcommand{\hatrk}{\ensuremath{\hat{r}_k}\xspace}
\newcommand{\dhatrk}{\ensuremath{\Delta\hatrk}\xspace}

\newcommand{\scale}{\ensuremath{\,\text{(scale)}}\xspace}
\newcommand{\fit}{\ensuremath{\,\text{(fit)}}\xspace}

\cmsNoteHeader{TOP-21-008}
\title{Measurement of the top quark pole mass using \texorpdfstring{\ttbarjet}{ttbar+jet} events in the dilepton final state in proton-proton collisions at \texorpdfstring{$\sqrt{s}=13\TeV$}{sqrt(s) = 13 TeV}}

\author*[cern]{The CMS Collaboration}

\date{\today}

\abstract{
    A measurement of the top quark pole mass \mtpole in events where a top quark-antiquark pair (\ttbar) is produced in association with at least one additional jet (\ttbarjet) is presented. This analysis is performed using proton-proton collision data at $\sqrt{s} = 13\TeV$ collected by the CMS experiment at the CERN LHC, corresponding to a total integrated luminosity of \lumivalueSix. Events with two opposite-sign leptons in the final state (\ee, \mumu, \emu) are analyzed. The reconstruction of the main observable and the event classification are optimized using multivariate analysis techniques based on machine learning. The production cross section is measured as a function of the inverse of the invariant mass of the \ttbarjet system at the parton level using a maximum likelihood unfolding. Given a reference parton distribution function (PDF), the top quark pole mass is extracted using the theoretical predictions at next-to-leading order. For the ABMP16NLO PDF, this results in $\mtpole=172.93\pm1.36\GeV$.
}

\hypersetup{%
pdfauthor={CMS Collaboration},%
pdftitle={Measurement of the top quark pole mass using ttbar+jet events in the dilepton final state in proton-proton collisions at sqrt(s) = 13 TeV},%
pdfsubject={CMS},%
pdfkeywords={CMS, top, dilepton, unfolding}}

\maketitle

\section{Introduction}
\label{sec:introduction}

The top quark is the most massive elementary particle known.
Its mass \mtop is a free parameter of the standard model (SM) Lagrangian and is an input to the global electroweak fits~\cite{bib:precision2,bib:precision3,bib:precision4,bib:precision5} and calculations of the Higgs boson self-coupling~\cite{bib:vacuum1,bib:vacuum2}.
The value of \mtop needs to be determined experimentally.
Direct measurements of \mtop at hadron colliders, being dominated by measurements in proton-proton (\pp) collisions at the CERN LHC, reach a precision on the order of 0.5\GeV~\cite{bib:directMass1,bib:directMass2,bib:directMass3,bib:directMass4,bib:directMass5,bib:directMass6,bib:directMass7,bib:topMasssingle}. The combined value of measurements by the CMS Collaboration at $\sqrt{s}=7$ and 8\TeV is $\mtop=172.44\pm0.48\GeV$~\cite{bib:directMass4}.
The direct measurements rely on the reconstruction of the top quark kinematic variables from its decay products and hence on the modeling provided by multipurpose Monte Carlo (MC) event generators.
These generators typically make use of parton shower models to approximate higher-order effects, while heuristic models tuned to the data are used to describe the color neutralization in the nonperturbative regime and the underlying event (UE).
This introduces ambiguities in the interpretation of the top quark mass assumed in the simulation, \mtopMC.
An additional intrinsic uncertainty on the order of about 0.5--1\GeV is expected when interpreting a direct measurement of \mtopMC in terms of the top quark pole mass, \mtpole~\cite{bib:massRelation1,bib:massRelation2,bib:massRelation3,bib:massRelation4,bib:massRelation5,bib:massRelation6,bib:massRelation7,bib:TillMatteo}.

Alternatively, the value of \mtop can be extracted by comparing cross section measurements to fixed-order theoretical predictions in well-defined renormalization schemes. The value of \mtpole was measured by both the ATLAS and CMS Collaborations~\cite{bib:poleMass1,bib:poleMass2,bib:ttbarMTMCCorr,bib:poleMass4,bib:poleMass5,bib:poleMass6,bib:sasha_multiDim}, reaching an uncertainty below 1\GeV in a recent CMS measurement~\cite{bib:sasha_multiDim} by using the multidifferential cross sections of top quark-antiquark (\ttbar) pair production.
The top quark mass defined in the modified minimal subtraction ($\overline{\text{MS}}$) scheme, referred to as the top quark running mass, was also measured~\cite{bib:runningMass1,bib:TillMatteo,bib:runningMass3} and its energy scale dependence was investigated in Ref.~\cite{bib:running}.

In this paper, the value of \mtpole is measured by using the normalized differential cross section of \ttbar production in association with at least one energetic jet (\ttbarjet)~\cite{bib:ttjet1}.
The measurement follows the approach developed in Refs.~\cite{bib:runningMass3,bib:ttjet1,bib:ttjet2}, where the normalized differential \ttbarjet cross section is measured as a function of the $\rho$ observable, defined as
\begin{linenomath}\begin{equation}
    \rho = \frac{2\mzero} {\mttbarjet},
\end{equation}\end{linenomath}
where \mttbarjet is the invariant mass of the \ttbarjet system.
The result of the measurement does not depend on the choice of the scaling constant \mzero, which is set to $\mzero=170\GeV$, as used in previous measurements. The sensitivity to \mtop is enhanced as compared to \ttbar production due to the presence of an additional jet because the kinematic parameters of the radiated gluons in \ttbarjet events depend on the mass of the top quark. High sensitivity to \mtpole is expected close to the production threshold, for $\rho>0.65$, while for high \mttbarjet, \eg, $\rho<0.55$, this sensitivity is small.
Such a measurement was performed by the ATLAS Collaboration using \pp collision data at $\sqrt{s}=7$ and 8\TeV~\cite{bib:poleMass4,bib:ttjetAtlas8}, corresponding to integrated luminosities of 4.6 and 20.2\fbinv, respectively, finding a value of $\mtpole = 171.1\,^{+1.2}_{-1.0}\GeV$ with the 8\TeV data set.

In this paper, a measurement of \mtpole is made by using the $\rho$ dependence at the center-of-mass energy of 13\TeV. The data were recorded by the CMS detector in 2016, corresponding to an integrated luminosity of \lumivalueSix.

In this analysis, the normalized differential cross section is measured at the detector level and is unfolded to the parton level. The unfolding is performed via the maximum likelihood method with profiled nuisance parameters for all systematic uncertainties, following the approach developed in Ref.~\cite{bib:running}.
The signal and background processes are fitted simultaneously, and the systematic uncertainties and their correlations are determined from data. This leads to significantly improved experimental precision over classical approaches where systematic uncertainties are not profiled~\cite{bib:TillMatteo}.

The paper is structured as follows.
After a brief description of the CMS experiment and the event reconstruction in Section~\ref{sec:CMSreco}, the data sets and simulation are presented in Section~\ref{sec:dataAndMCSamples}, together with the definition of the signal.
The event selection is explained in Section~\ref{sec:eventSelection}.
In Section~\ref{sec:dnns}, a detailed description of multivariate analysis techniques developed for this measurement is given.
The maximum likelihood fit for the signal extraction and unfolding is described in Section~\ref{sec:fit}, followed by a discussion of the systematic uncertainties in Section~\ref{sec:systematics}.
The determination of \mtpole is presented in Section~\ref{sec:mass}, and a summary is given in Section~\ref{sec:summary}.
Tabulated results are provided in the HEPData record for this analysis~\cite{hepdata}.

\section{The CMS detector and event reconstruction}
\label{sec:CMSreco}
The central feature of the CMS apparatus is a superconducting solenoid of 6\unit{m} internal diameter, providing a magnetic field of 3.8\unit{T}. Within the solenoid volume are a silicon pixel and strip tracker, a lead tungstate crystal electromagnetic calorimeter (ECAL), and a brass and scintillator hadron calorimeter (HCAL), each composed of a barrel and two endcap sections. Forward calorimeters extend the pseudorapidity (\sigeta) coverage provided by the barrel and endcap detectors. Muons are detected in gas-ionization chambers embedded in the steel flux-return yoke outside the solenoid.
Events of interest are selected using a two-tiered trigger system. The first level (L1), composed of custom hardware processors, uses information from the calorimeters and muon detectors to select events at a rate of around 100\unit{kHz} within a fixed latency of about 4\mus~\cite{Sirunyan:2020zal}. The second level, known as the high-level trigger (HLT), consists of a farm of processors running a version of the full event reconstruction software optimized for fast processing, and reduces the event rate to around 1\unit{kHz} before data storage~\cite{Khachatryan:2016bia}.
A more detailed description of the CMS detector, together with a definition of the coordinate system used and the relevant kinematic variables, can be found in Ref.~\cite{Chatrchyan:2008zzk}.

The primary vertex (PV) is taken to be the vertex corresponding to the hardest scattering in the event, evaluated using tracking information alone, as described in Section 9.4.1 of Ref.~\cite{CMS-TDR-15-02}.

The particle-flow (PF) algorithm~\cite{CMS-PRF-14-001} aims to reconstruct and identify each individual particle in an event, with an optimized combination of information from the various elements of the CMS detector. The energy of photons is obtained from the ECAL measurement. The energy of electrons is determined from a combination of the electron momentum at the PV as determined by the tracker, the energy of the corresponding ECAL cluster, and the energy sum of all bremsstrahlung photons spatially compatible with originating from the electron track. The energy of muons is obtained from the curvature of the corresponding track. The energy of charged hadrons is determined from a combination of their momentum measured in the tracker and the matching ECAL and HCAL energy deposits, corrected for the response function of the calorimeters to hadronic showers. Finally, the energy of neutral hadrons is obtained from the corresponding corrected ECAL and HCAL energies.

For each event, hadronic jets are clustered from PF particles using the infrared- and collinear-safe anti-\kt algorithm with a distance parameter of 0.4. Jet momentum is determined as the vectorial sum of all particle momenta in the jet, and is found from simulation to be, on average, within 5 to 10\% of the true momentum over the whole transverse momentum (\pt) spectrum and detector acceptance. Additional \pp interactions within the same or nearby bunch crossings (pileup) can contribute additional tracks and calorimetric energy depositions, increasing the apparent jet momentum. To mitigate this effect, tracks identified to be originating from pileup vertices are discarded and an offset correction is applied to correct for remaining contributions. Jet energy corrections are derived from simulation studies so that the average measured energy of jets becomes identical to that of particle-level jets. In situ measurements of the momentum balance in dijet, photon+jet, {\PZ}+jet, and multijet events are used to determine any residual differences between the jet energy scale in data and in simulation, and appropriate corrections are made~\cite{Khachatryan:2016kdb}. No dedicated flavor-dependent corrections are made. Additional selection criteria are applied to each jet to remove jets potentially dominated by instrumental effects or reconstruction failures. The jet energy resolution amounts typically to 15--20\% at 30\GeV, 10\% at 100\GeV, and 5\% at 1\TeV, and corrections as determined from QCD dijet and $\gamma$+jet events~\cite{Khachatryan:2016kdb} are applied.

The missing transverse momentum vector \ptvecmiss is computed as the negative vector sum of the \ptvec of all the PF candidates in an event, and its magnitude is denoted as \ptmiss~\cite{Sirunyan:2019kia}. The value of \ptvecmiss is modified to account for corrections to the energy scale of the reconstructed jets in the event.
A pileup per particle identification algorithm~\cite{Bertolini:2014bba, bib:pileupjetid} is applied to reduce the pileup dependence of \ptvecmiss and the PF candidates are weighted by their probability to originate from the PV~\cite{Sirunyan:2019kia}.

Electrons are measured in the range $\abseta < 2.5$, and the single-electron trigger efficiency is about 80\%. The electron momenta are estimated by combining energy measurements in the ECAL with momentum measurements in the tracker~\cite{CMS-DP-2020-021}. The efficiency to reconstruct and identify electrons is better than 95\%~\cite{bib:egamma}.
The momentum resolution for electrons with $\pt \approx 45\GeV$ from $\PZ \to \Pe \Pe$ decays ranges from 1.6 to 5\%. It is generally better in the barrel region ($\abseta < 1.479$) than in the endcaps ($1.479 < \abseta < 3.0$), and also depends on the bremsstrahlung energy emitted by the electron as it traverses the material in front of the ECAL~\cite{bib:egamma,CMS-DP-2020-021}.

Muons are measured in the range $\abseta < 2.4$, with detection planes made using three technologies: drift tubes, cathode strip chambers, and resistive plate chambers. The single-muon trigger efficiency exceeds 90\% over the full \sigeta range, and the efficiency to reconstruct and identify muons is greater than 96\%. Matching muons to tracks measured in the silicon tracker results in a relative \pt resolution, for muons with \pt up to 100\GeV, of 1\% in the barrel and 3\% in the endcaps. The \pt resolution in the barrel is better than 7\% for muons with \pt up to 1\TeV~\cite{Sirunyan:2018}.

\section{Data samples and event simulation}
\label{sec:dataAndMCSamples}

The analysis uses \pp collision data recorded by the CMS experiment in 2016, corresponding to an integrated luminosity of \lumivalueSix. The events are required to pass one of several HLT selection criteria, including the presence of either one or two leptons (electrons or muons). The single-electron (muon) trigger requires a minimum \pt of 27 (24)\GeV.
In the case of the dielectron triggers, at least two lepton candidates with a minimum \pt of 23 and 12\GeV are required, while for the dimuon triggers the respective thresholds are 17 and 8\GeV.
For the electron-muon triggers, the selection requirements are $\pt>23\GeV$ and $\pt>8\GeV$ ($\pt>13\GeV$) if the muon (electron) is the lepton with higher \pt.

Signal and background \ttbar processes are simulated using the \POWHEG (version 2)~\cite{bib:powhegV2_1,bib:powhegV2_2,bib:powhegV2_3} MC generator at next-to-leading order (NLO) in quantum chromodynamics (QCD) for the matrix element (ME) calculations. The top quark mass in the simulation is fixed to $\mtopMC=172.5\GeV$ and the proton structure is described using the NNPDF~3.1 PDF at next-to-NLO (NNLO)~\cite{bib:NNPDF1,bib:NNPDF2}. Parton showering is performed using \PYTHIA (version 8.230)~\cite{bib:pythia8} with the CP5 tune~\cite{bib:cp5}. The \hdamp parameter in \POWHEG, which effectively regulates the matching scale to the parton shower, is set to $\hdamp=1.379\mtopMC$~\cite{bib:cp5}. The \ttbar simulation is split into different components, modeling the signal and the background contributions, as described in Section~\ref{subsec:partonlevel}.

Additionally, simulated samples of the \ttbar process are generated with up to two extra partons at the NLO ME level using \MGvATNLO~\cite{bib:amcatnlo}. These are used to derive the calibrations of the neural networks (NNs), as discussed in Section~\ref{sec:dnns}. Events are matched using the \FXFX~\cite{bib:fxfx} prescription to the \PYTHIA parton shower, while \textsc{madspin}~\cite{bib:madspin} is used to model the decays of the top quark and antiquark at leading order (LO). The matching scale for the \MGvATNLO[{\FXFX}] sample is set to 40\GeV.

The background contributions arise from single top quark production in association with a \PW boson (\tW), $\PZ/\PGg^\ast$ or \PW bosons produced with additional jets (\Zjets, \Wjets), and diboson production ($\PW\PW$, $\PW\PZ$, $\PZ\PZ$). The contributions from the diboson and \Wjets production processes are labeled as ``Other'' in the figures.
For all the background samples, \PYTHIA is used to model the parton showering, hadronization, and multiparton interactions. The \tW process is simulated using \POWHEG~\cite{bib:powhegV1_1,bib:powhegV1_2} at NLO with the same tune and modeling parameters as used for the \ttbar process. The CUETP8M1~\cite{bib:hdamp,bib:m2t4,bib:m2t4_2} tune is used for the background simulation. The \Zjets process is simulated using \MGvATNLO at LO with up to four additional partons at ME level, and is matched to the \PYTHIA parton shower using the \textsc{mlm}~\cite{bib:mlm,bib:mlm2} prescription. Similarly, \Wjets samples are produced using \MGvATNLO at LO with up to four additional partons at ME level. To derive the NN calibrations, an independent \Zjets simulation is generated with \MGvATNLO at NLO using the \FXFX merging prescription. Diboson production is simulated at LO with \PYTHIA.

{\tolerance=1100
All predictions are normalized to their theoretical cross section and the corresponding integrated luminosity of the data set.
For the \ttbar signal, the value of the cross section is $830.91\,^{+20.39}_{-29.96}\scale\,^{+3.92}_{-3.83}\,(\text{PDF+}\alpS)\unit{pb}$, as calculated with the \textsc{Top++} (version 2.0) program~\cite{bib:top++} at NNLO precision.
The calculation assumes $\mtop=172.5\GeV$, makes use of the NNPDF3.1 PDF set at NNLO, and includes the resummation of next-to-next-to-leading-logarithmic soft-gluon terms~\cite{bib:xsec_ttbar1,bib:xsec_ttbar2,bib:xsec_ttbar3,bib:xsec_ttbar4,bib:xsec_ttbar5,bib:xsec_ttbar6}.
The labels ``scale'' and \alpS refer to the uncertainty assessed by varying the renormalization (\muR) and factorization (\muF) scales and the value for the strong coupling constant, respectively.
The remaining cross sections are calculated at approximate NNLO for \tW production~\cite{bib:xsec_tW}, NLO for \Zjets and diboson production~\cite{bib:xsec_diboson}, and NNLO for \Wjets~\cite{bib:xsec_wjets} production.
\par}

For all simulated samples, the CMS detector response is simulated with \GEANTfour~\cite{bib:geant4}. To model the effect of pileup, additional minimum-bias interactions are added to the simulated events. Weights are used with the simulated events to reproduce the pileup distribution in the data. The weights are estimated from the measured bunch-to-bunch instantaneous luminosity, assuming a total inelastic \pp cross section of 69.2\unit{mb}~\cite{bib:minbias}.

\subsection{Signal definition at the parton level}
\label{subsec:partonlevel}
The simulated signal consists of \ttbar events in which a dilepton and at least one jet emitted at the parton level with $\pt>30\GeV$ and $\abseta<2.4$ are produced.
The parton level is defined after parton showering with on-shell top quarks.
Additional jets are clustered by applying the anti-\kt algorithm~\cite{Cacciari:2008gp, Cacciari:2011ma} with a distance parameter of 0.4, using all particles before hadronization that do not stem from the top quark decays.
The \ttbarjet signal is obtained from the nominal \ttbar simulation by selecting events with at least one additional parton-level jet passing the selection described above. The remaining events are considered as background and labeled as \ttbarnojet.
The \ttbarjet signal is further split into subprocesses that are defined based on the value of $\rho$ at parton level in the following binning: 0--0.3, 0.3--0.45, 0.45--0.7, and 0.7--1. The binning choice is explained in Section~\ref{sec:fit}.
The simulated \ttbar signal demands at least two leptons, including those from the decays of tau leptons. The remaining \ttbar events are considered as background.
The unfolded results are extrapolated to the full phase space using the \ttbar MC simulation.

\section{Event selection}
\label{sec:eventSelection}

Data events containing two opposite-sign leptons (\ee, \mumu, \emu) are selected. To suppress events from low-mass resonance decays and the Drell-Yan process, the dilepton invariant mass (\mll) is required to be larger than 20\GeV. The \Zjets background contribution is reduced by discarding events with $\ptmiss<40\GeV$ and \mll in the \PZ mass window between 76 and 106\GeV, for the same-flavor lepton channels.

The leading (subleading) electron candidate is required to have $\pt>25$ (20)\GeV and $\abseta<2.4$. Electrons reconstructed in the ECAL barrel-endcap transition region ($1.44<\abseta<1.57$) are rejected. A relative isolation parameter (\Isorel) is calculated for each electron candidate, defined as the ratio of the \pt sum of neutral and charged hadron and photon PF candidates,
within a distance of $\DR=\sqrt{\smash[b]{(\Delta\eta)^2+(\Delta\phi)^2}}=0.3$ in \sigeta-$\phi$ space, to the \pt of the electron candidate, where $\phi$ is the azimuthal angle.
Depending on the electron \pt, the maximum allowed value for \Isorel varies from 0.05 to 0.1. Additional corrections are made to remove contributions from pileup~\cite{bib:egamma}. To reject misidentified electron candidates and photon conversions, additional identification requirements are imposed~\cite{bib:egamma}.

Muon candidates are selected with the same \pt and \sigeta requirements as for the electrons, except for removing the transition-region requirement.
An isolation requirement of $\Isorel<0.15$ is applied, where all particle momenta within a distance of $\DR<0.4$ from the muon are considered and pileup effects are taken into account.
Muon candidates are further identified by their specific signature in the detector~\cite{Sirunyan:2018}.

Jets with $\pt>30\GeV$ and $\abseta<2.4$ are selected, and the value of {\DR} between the jet and the selected lepton candidates must be greater than 0.4.
To further reject jets arising from pileup interactions, a dedicated pileup jet identification algorithm~\cite{bib:pileupjetid} is employed for jets with \pt between 30 and 50\GeV.
Jets originating from the hadronization of \PQb quarks (called \PQb jets) are identified (tagged) using the \textsc{DeepCSV} algorithm~\cite{bib:btv}, which makes use of associated track and secondary-vertex information as inputs to a deep NN.
The chosen operating point for the discriminator has an efficiency of about 80--90\% for \PQb jets, for misidentification (``mistag'') rates of 10 and 40\% for light-quark and \PQc quark jets, respectively~\cite{bib:btv}.
The energy measurement of \PQb jets is improved by about 6--12\% using a deep NN estimator~\cite{bib:bregression}.

Lepton trigger and identification efficiencies, lepton momentum resolutions, and \PQb tagging efficiencies are corrected in simulation to match the values in data.

The typical purities for the \ttbarjet signal obtained without requirements on the jet or \PQb jet multiplicity range from about 15\% in the \ee and \mumu channels to about 30\% in the \emu channel. When selecting only events with at least three reconstructed jets, of which one is \PQb~tagged, the purity is around 75\%.

\begin{figure}[!tp]
    \centering
        \includegraphics[width=.49\textwidth]{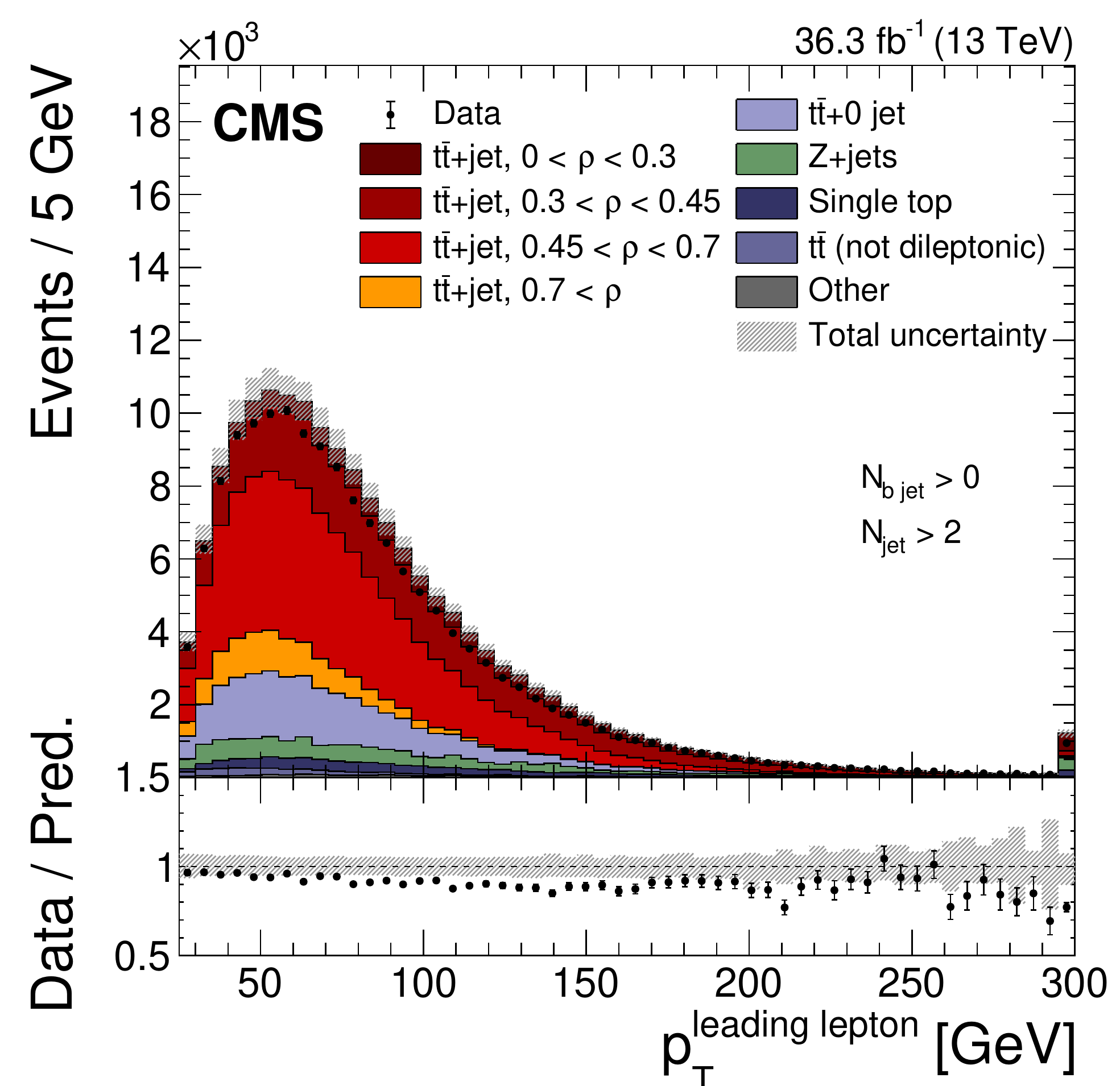}
        \hfill
        \includegraphics[width=.49\textwidth]{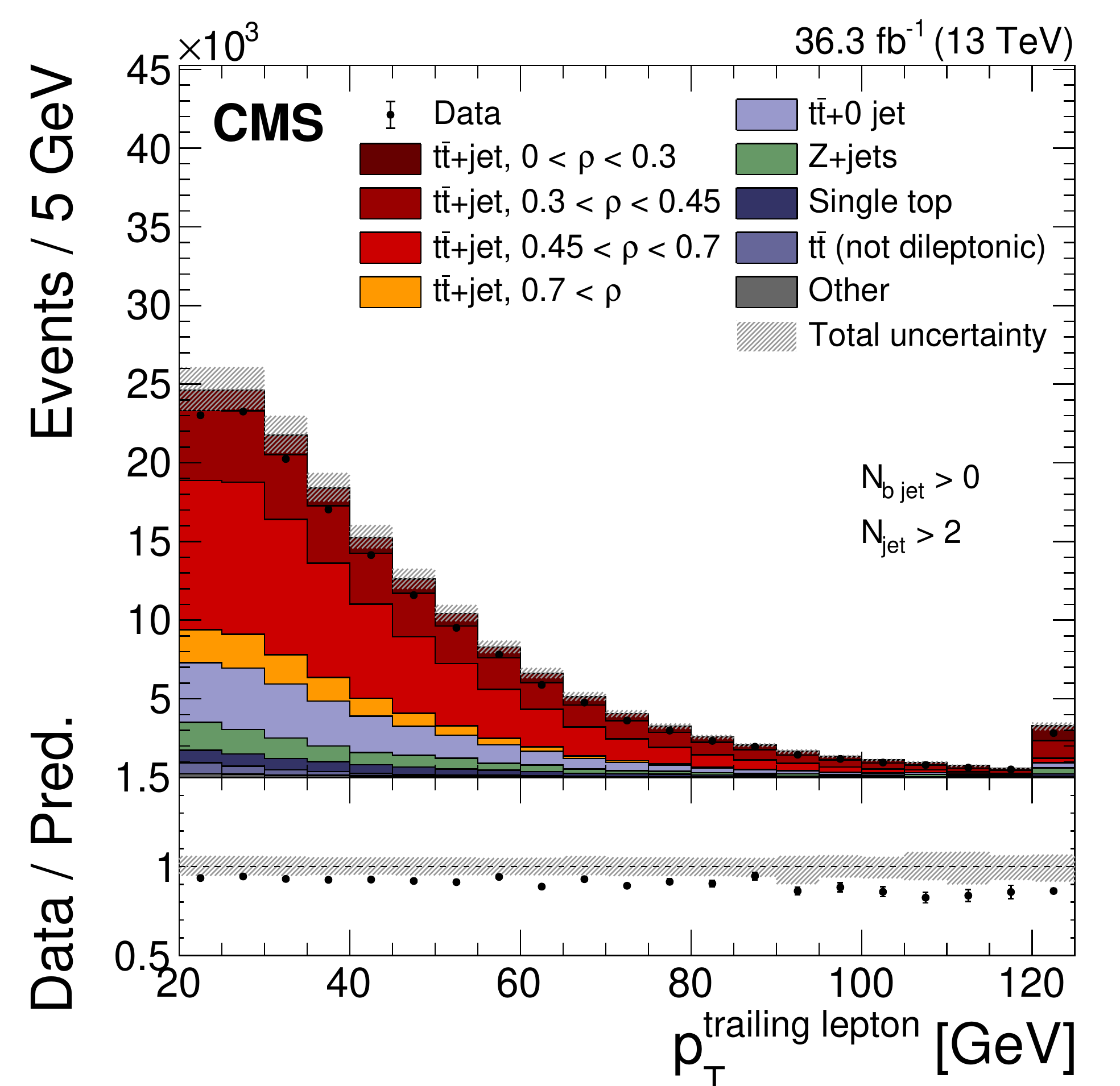}\\
        \includegraphics[width=.49\textwidth]{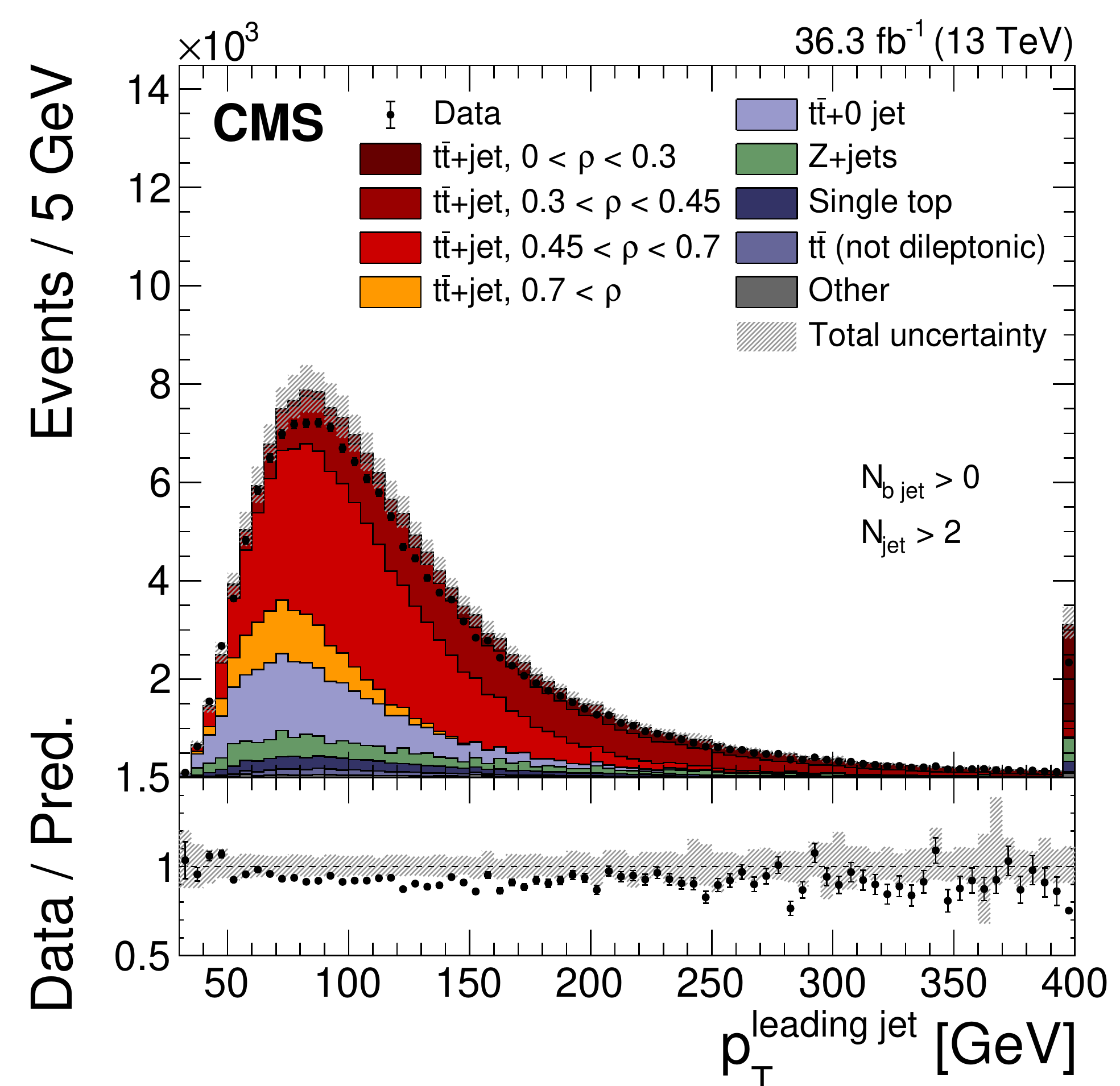}
        \hfill
        \includegraphics[width=.49\textwidth]{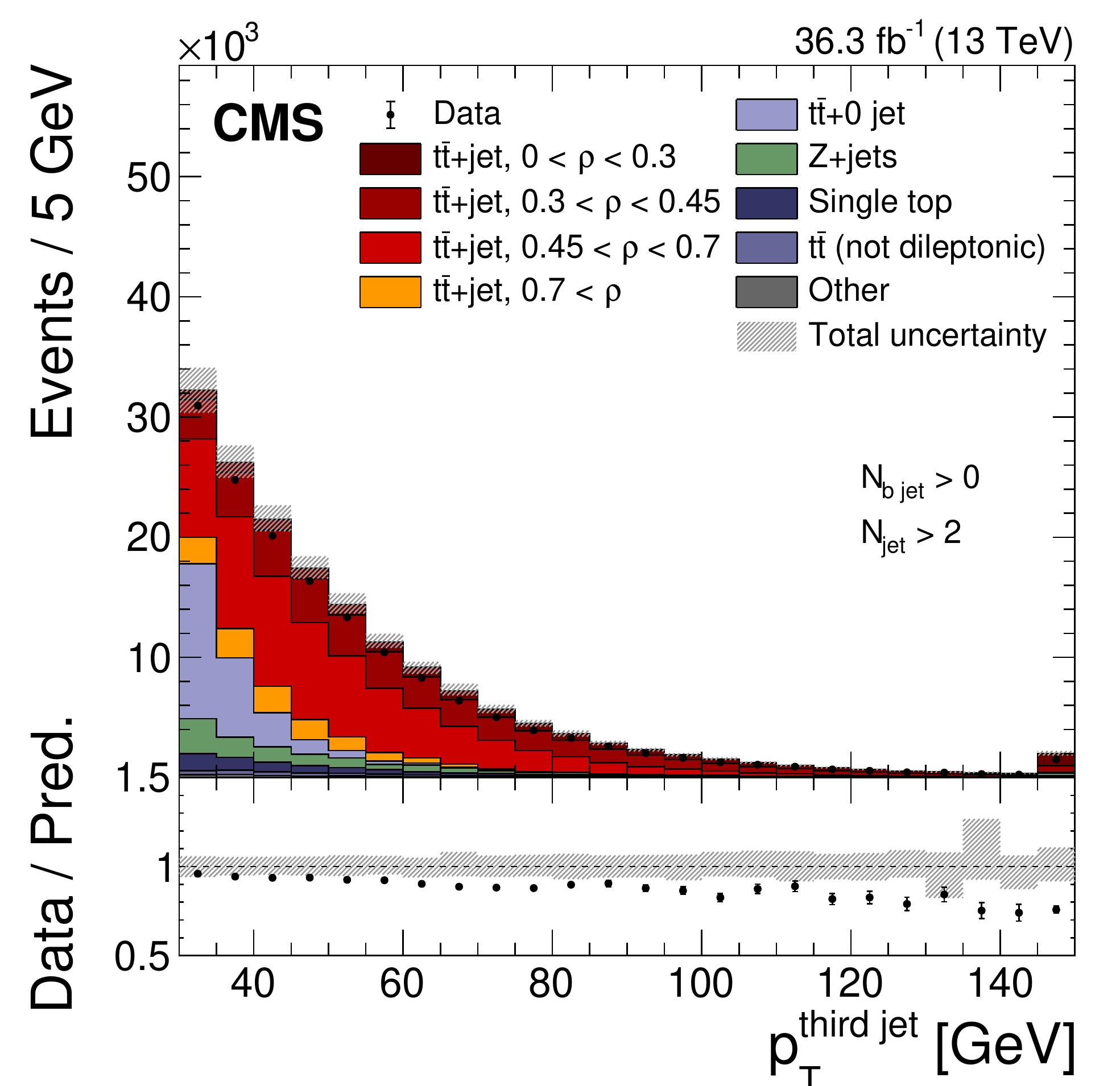}
    \caption{The observed (points) and predicted (stacked histograms) signal and background yields as a function of the leading (upper left) and subleading (upper right) lepton \pt and leading (lower left) and third-highest (lower right) jet \pt after applying the signal selection. The vertical bars on the points represent the statistical uncertainty in the data. The hatched band represents the total uncertainty in the sum of the simulated signal and background predictions. The lower panels show the ratio of the data to the sum of the signal and background predictions.}
    \label{fig:controlplots1}
\end{figure}

\begin{figure}[!tp]
    \centering
        \includegraphics[width=.49\textwidth]{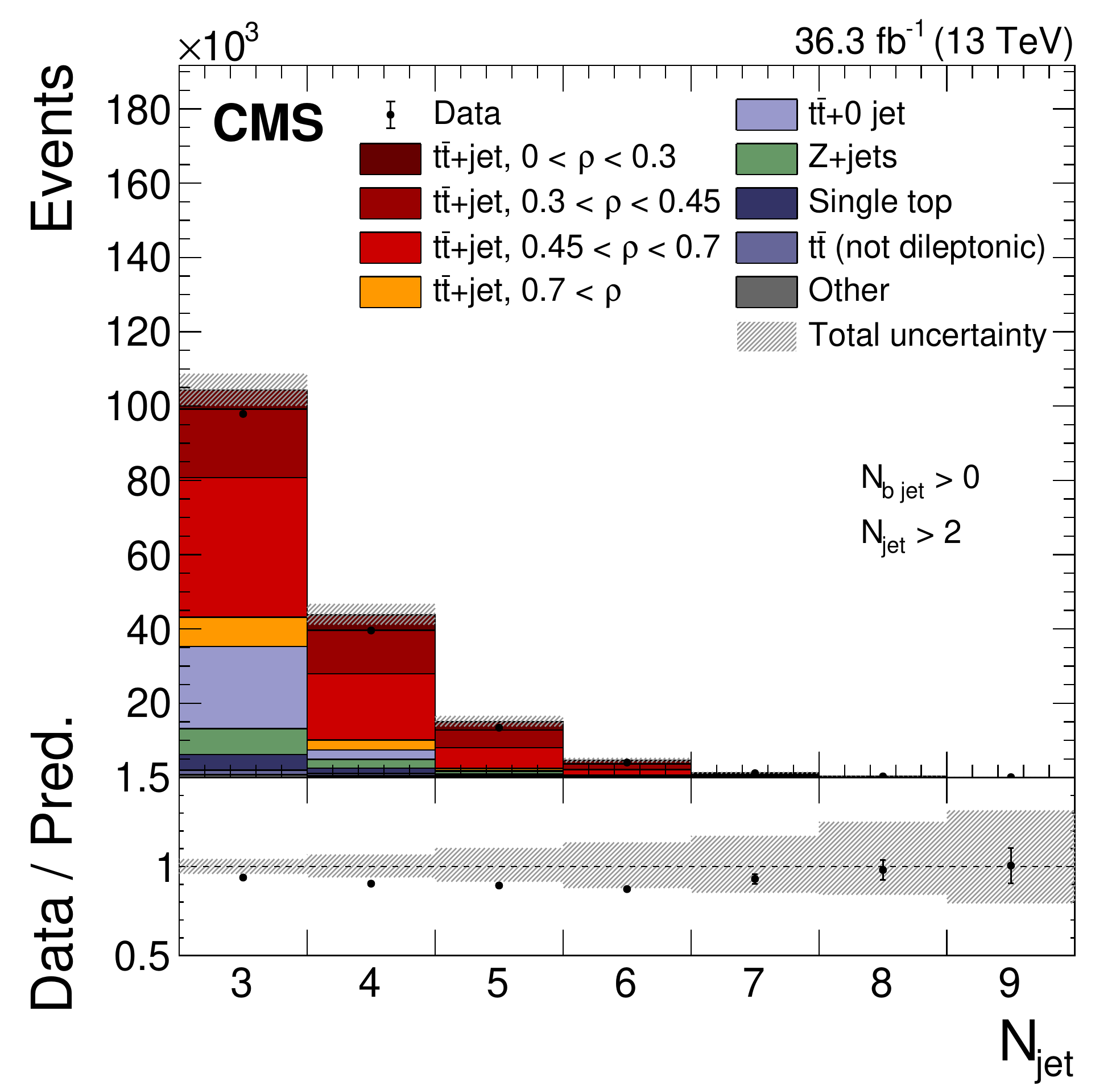}
        \hfill
        \includegraphics[width=.49\textwidth]{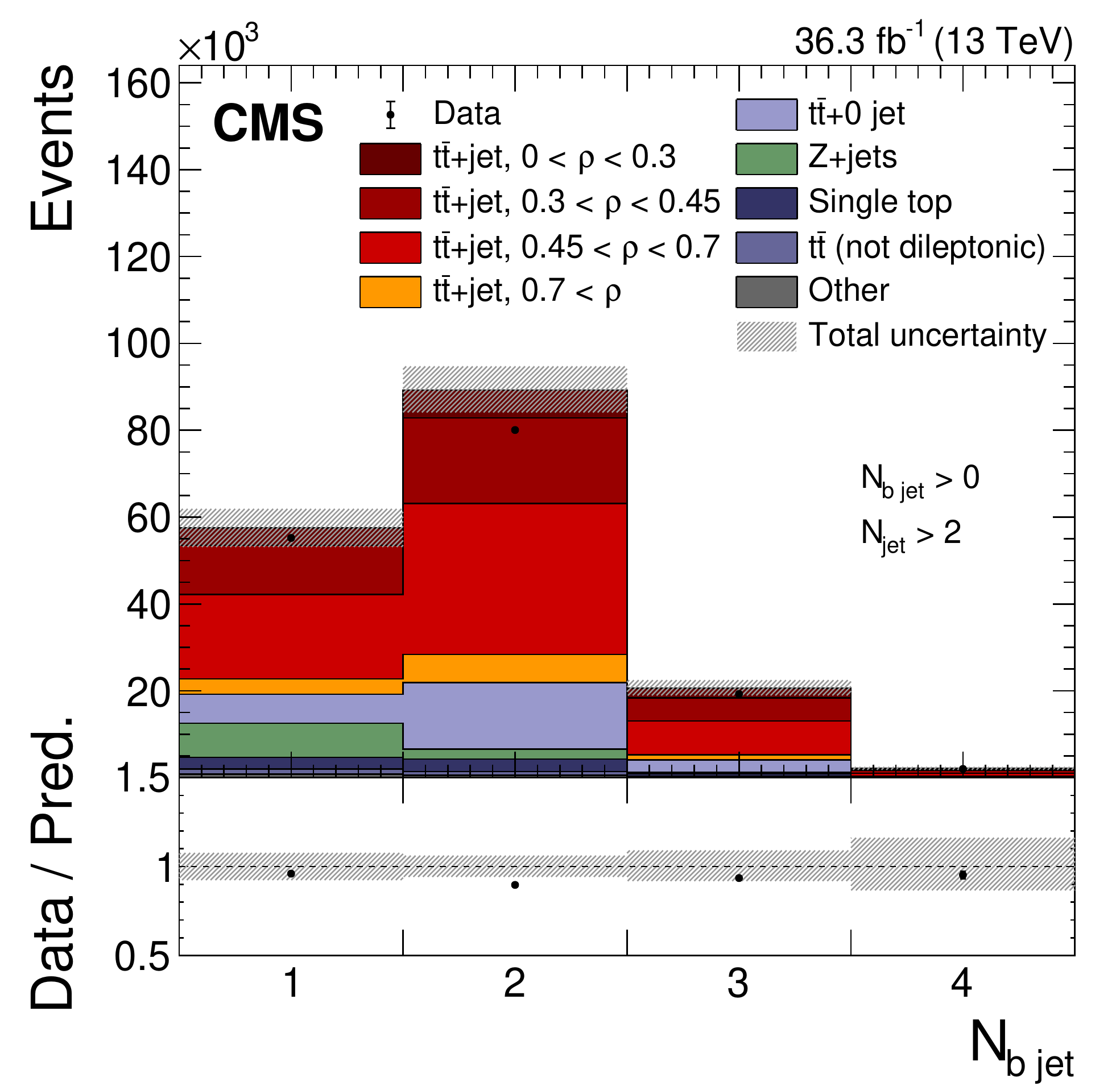}
    \caption{The observed (points) and predicted (stacked histograms) signal and background yields as a function of the jet (left), and \PQb jet (right) multiplicities, after applying the signal selection. The vertical bars on the points represent the statistical uncertainty in the data. The hatched band represents the total uncertainty in the sum of the simulated signal and background predictions. The lower panels show the ratio of the data to the sum of the signal and background predictions.}
    \label{fig:controlplots2}
\end{figure}

Figures~\ref{fig:controlplots1} and \ref{fig:controlplots2} show the distributions of several kinematic observables from data and simulation for the combined dilepton candidates.
In general, the MC simulation describes the data well.
A small discrepancy between the normalization of the data and simulation is observed in several of the distributions.
However, this does not affect this measurement since the normalization is determined from the likelihood fit, as described in Section~\ref{sec:fit}, and the normalized cross section is used to extract \mtpole.
Some trends can be observed in the jet and lepton momentum distributions, where the spectrum predicted by the simulation has a higher average value than the one observed in data.
The effect is related to the modeling of the top quark \pt in the \POWHEGPYTHIAEight\ \ttbar simulation, and similar observations have been made in previous measurements~\cite{bib:kinreco,bib:ttbar1,bib:ttbar2,bib:ttbar3} when comparing the measured distributions to detector- or particle-level predictions.
The jet multiplicity and \PQb jet multiplicities from data are well described by the simulation within the total uncertainties.

\section{Multivariate analysis}
\label{sec:dnns}
To reconstruct $\rho$ as the variable of interest, kinematic reconstruction techniques must be employed.
These methods typically rely on imposing constraints to reconstruct the longitudinal components of the momenta of the two neutrinos from the \PW boson leptonic decays in \ttbar events containing dileptons. Several methods have been established and used in previous differential \ttbar cross section measurements~\cite{bib:Sonnenschein,bib:kinreco,bib:sasha_multiDim,bib:ttbar1,bib:ttbar2,bib:ttbar3,bib:running}.

Two multivariate analysis (MVA) approaches are developed to mitigate constraints on the kinematic reconstruction and and improve the signal-versus-background discrimination.
The MVA methods developed for this analysis are described below.

\subsection{Reconstruction of top quark kinematic variables}
\label{subsec:kinreco}
Two analytical kinematic reconstruction approaches referred to as ``full kinematic reconstruction'' and ``loose kinematic reconstruction'' are used. Solutions of both methods are considered as inputs to the MVAs.
For the full kinematic reconstruction, the four-momenta of the top quark and antiquark are determined using the following constraints: total \ptvec conservation; the mass of the \PW boson (80.4\GeV), and the masses of the top quark and antiquark (172.5\GeV). Reconstructed \ptvecmiss in the event is assumed to originate from the two neutrinos produced in the two \PW boson decays. Possible ambiguities in the algebraic solutions are solved by taking the solution with the smallest \ttbar invariant mass. To increase the reconstruction efficiency and account for the detector resolution, the reconstruction procedure is repeated 100 times for each event. Each time, reconstructed momenta of the measured jets and leptons are randomly smeared according to their resolution, where the smearing effect is propagated to \ptvecmiss. Each of the 100 solutions is weighted based on the true distribution of the invariant mass of a lepton and a \PQb jet (\mlb) stemming from the top quark decays as given by the simulation, where the weights for both decay chains of the \PQt and \PAQt are multiplied. All combinations of jets and leptons with $\mlb<180\GeV$ are considered and ranked by the presence of \PQb jets in the solutions, where solutions with a higher \PQb jet multiplicity are preferred. Further details about the reconstruction method can be found in Ref.~\cite{bib:kinreco}.

In the loose kinematic reconstruction~\cite{bib:sasha_multiDim}, only the \ttbar system as a whole is reconstructed. The advantage of this method is that the \mtop constraint is not used, which makes this method especially useful for the determination of \mtop. Similarly to the full kinematic reconstruction, only jet--lepton combinations with $\mlb<180\GeV$ are considered.
Combinations based on the different assignments are ranked by the number of \PQb jets, and solutions with more \PQb jets are preferred. Among combinations with the same number of \PQb jets, the one with the highest \pt jets are chosen.
The kinematic variables of the \NN system are obtained as follows: $\ptvec^{\,\NN}$ is set equal to \ptvecmiss and its longitudinal component and energy are set equal to that of the charged-lepton-pair system.
Furthermore, constraints are imposed on the reconstructed invariant masses of the \WW, $m(\WW)>2m_{\PW}$, and neutrino pair, $m(\NN)>0$.

To optimize the resolution of the $\rho$ observable, an MVA approach is developed, based on a regression NN, where the target variable is the parton-level $\rho$ as defined in the MC.
Using the \textsc{TensorFlow}~\cite{bib:tensorflow} package and the \textsc{Keras}~\cite{bib:keras} backend, a fully connected feed-forward NN is trained. An NN with two hidden layers of 512 nodes each is used. Low-level variables include basic event information such as the \pt of a reconstructed particle, whereas high-level variables include such things as the solutions to the kinematic reconstruction. Starting from a set of more than 100 low- and high-level input variables, the ten most relevant input variables are identified and selected, while the remaining ones are discarded. All events passing the selection described in Section~\ref{sec:eventSelection} and having at least three reconstructed jets are considered in the training.
As introduced in Section~\ref{sec:dataAndMCSamples}, a statistically independent training sample is obtained using \ttbar MC events simulated with the \MGvATNLO[{\FXFX}] event generator interfaced with \PYTHIAEight. All NN hyperparameters are optimized using a Bayesian optimizer approach~\cite{bib:bayesian1,bib:bayesian2,bib:bayesian3}, and the final NN response is cross-checked using the \POWHEGPYTHIAEight simulation.
Input variables, ordered by their impact, used for the regression NN are
\begin{itemize}
    \item solution for $\rho$ using the loose kinematic reconstruction,
    \item solution for $\rho$ using the full kinematic reconstruction,
    \item invariant mass of the dilepton and subleading jet system,
    \item invariant mass of the leading lepton and subleading jet system,
    \item \pt of the subleading lepton,
    \item invariant mass of the dilepton,
    \item invariant mass of the subleading lepton and subleading jet system,
    \item invariant mass of the subleading lepton and leading jet system,
    \item invariant mass of the dilepton and leading jet system,
    \item \ptmiss.
\end{itemize}
If no solution from the kinematic reconstruction is obtained, a value of zero is used instead.
The agreement between data and simulation for all inputs is evaluated using goodness-of-fit tests based on a saturated model~\cite{bib:gof}.
All the input variables are found to have a $p$-value${} > 0.08$, when compared between data and simulation.
The test procedure and the final result are independent of the cross section and not sensitive to \mtop.

\begin{figure}[!ht]
    \centering
        \includegraphics[width=.49\textwidth]{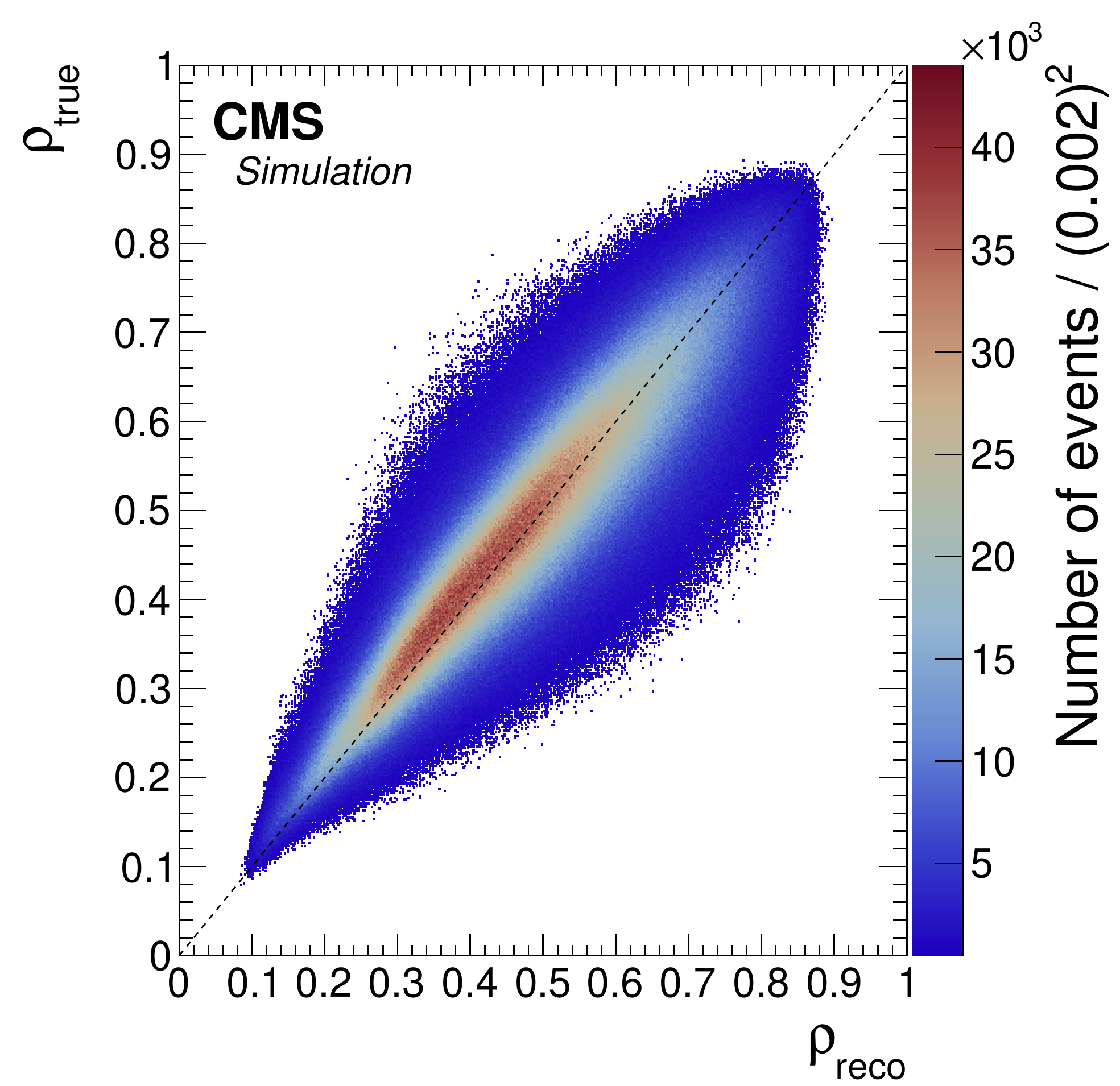}
        \hfill
        \includegraphics[width=.49\textwidth]{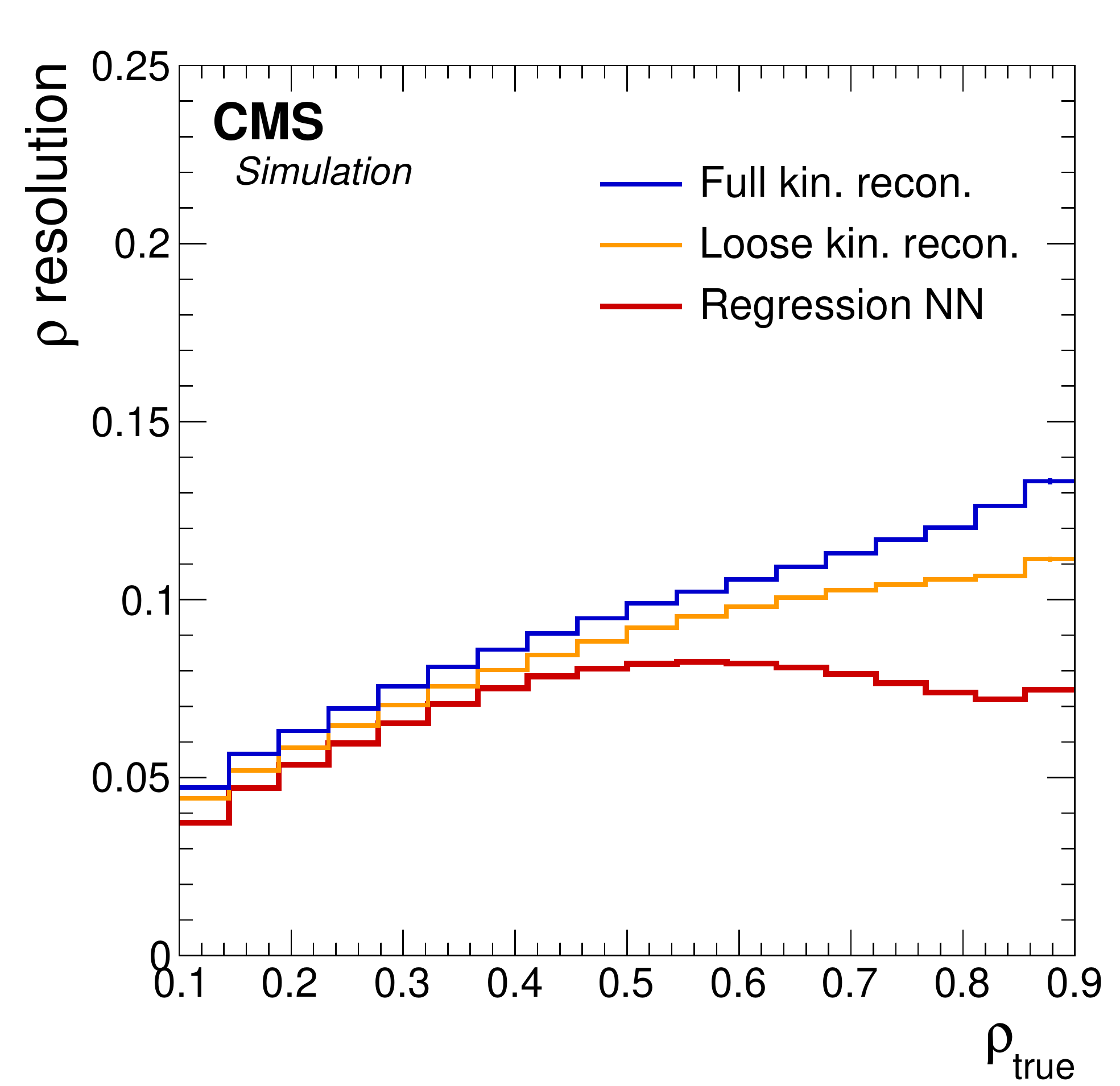}
    \caption{The correlation between \rhotrue and \rhoreco is shown for the regression NN reconstruction method (left). The \rhoreco resolution, defined in the text, as a function of \rhotrue (right) for the full (blue line) and loose (orange line) kinematic reconstructions and the regression NN (red line) methods. The number of events per bin in the left plot is shown by the color scale.}
    \label{fig:rho_performance_comp}
\end{figure}

\begin{figure}[!ht]
    \centering
        \includegraphics[width=.49\textwidth]{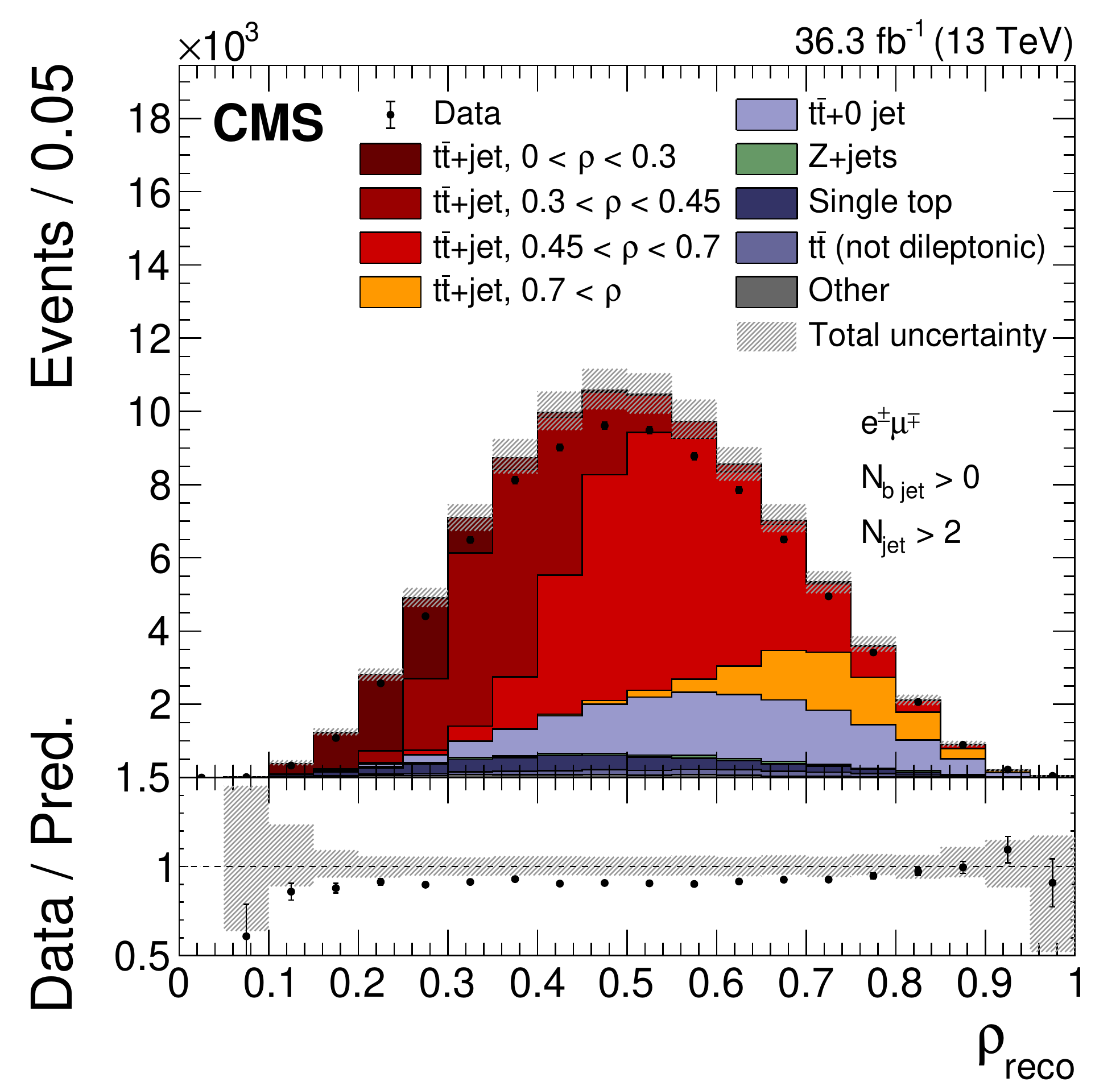}
        \hfill
        \includegraphics[width=.49\textwidth]{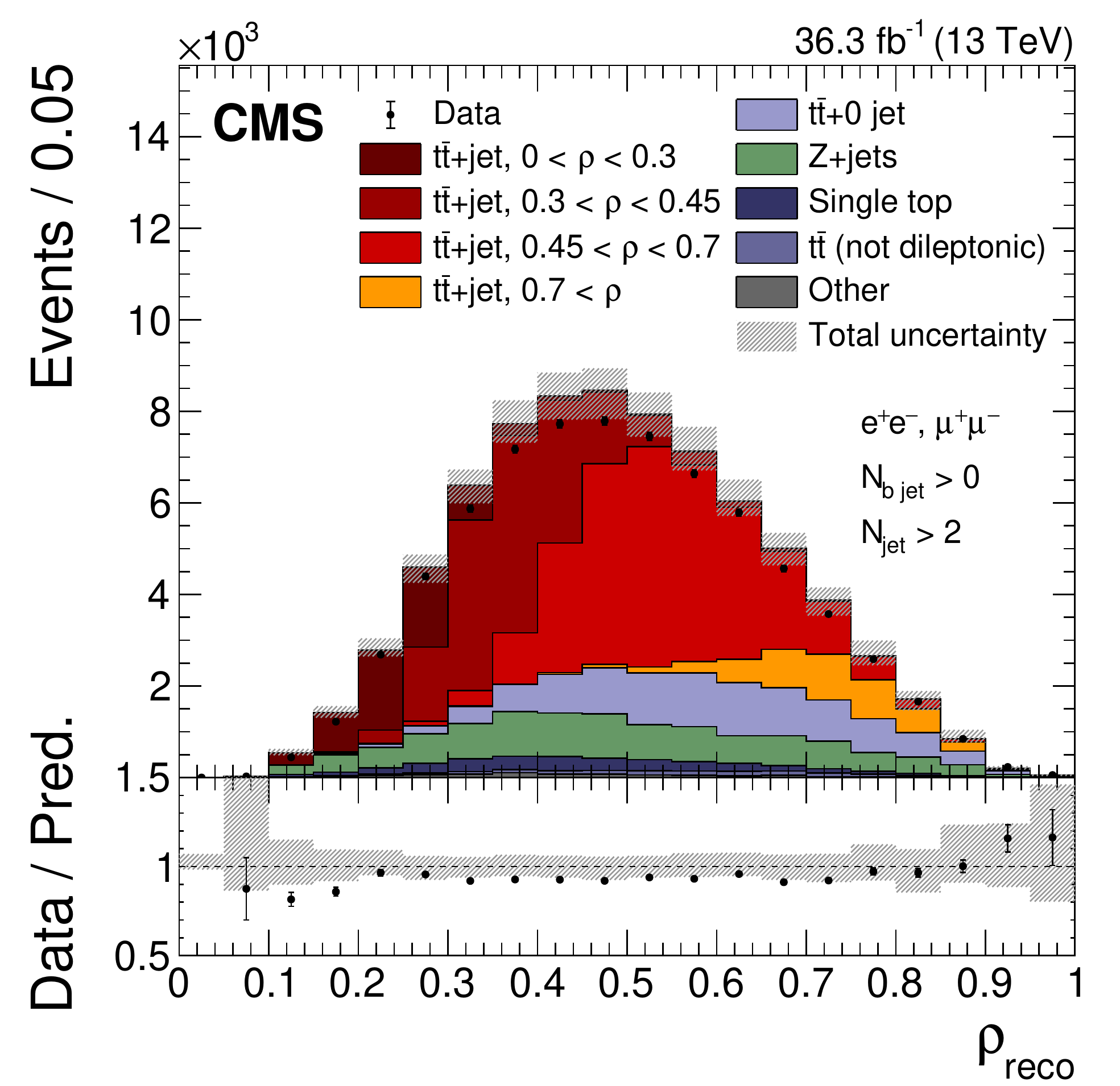}
    \caption{The observed (points) and MC predicted (stacked histograms) signal and background yields as a function of \rhoreco as determined by the NN reconstruction method for the \emu (left) and same-flavor dilepton channels (right). The vertical bars on the points represent the statistical uncertainty in the data. The hatched band represents the total uncertainty in the sum of the simulated signal and background predictions. The lower panels show the ratio of the data to the sum of the signal and background predictions.}
    \label{fig:rho_shapes}
\end{figure}

The performance of the NN regression is shown in Fig.~\ref{fig:rho_performance_comp}, where the correlation between the parton-level $\rho$ value (\rhotrue) and the reconstructed value (\rhoreco) is shown. The regression correlation coefficient is 0.87, compared to 0.78 (0.84) for the loose (full) kinematic reconstruction. In the same figure, the \rhoreco resolution of the obtained solutions is displayed as a function of \rhotrue.
The \rhoreco resolution in each bin of \rhotrue is defined as the root-mean-square of the difference between \rhoreco and \rhotrue, divided by $1+\langle\rhotrue - \rhoreco\rangle$. A resolution between 0.05--0.08 is obtained over the entire spectrum, which is a significant improvement compared to the two analytical reconstruction methods that have a resolution in the range 0.10--0.13 for $\rhotrue>0.5$.
Moreover, unlike the kinematic reconstruction methods, which are affected by reconstruction inefficiencies, the NN approach gives a measurement of $\rho$ for every event. In this analysis, the solution of the MVA regression is used, and is indicated with \rhoreco. Distributions of \rhoreco for the data and simulation are shown in Fig.~\ref{fig:rho_shapes} for the \emu (left plot) and same-flavor (right plot) dilepton channels.

\subsection{Event classification}
\label{subsec:classification}
\begin{figure}[!ht]
    \centering
        \includegraphics[width=.49\textwidth]{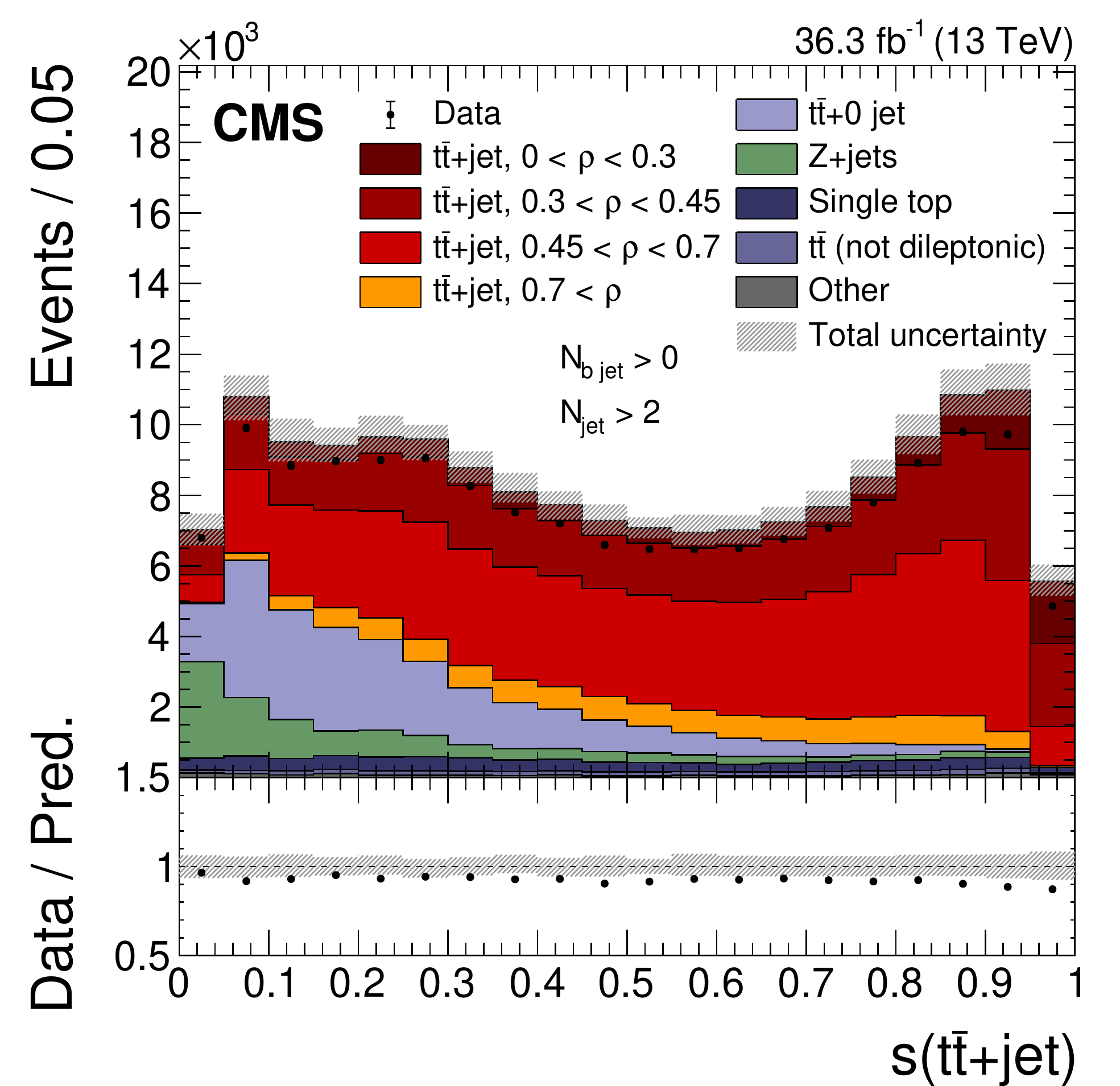}
        \hfill
        \includegraphics[width=.49\textwidth]{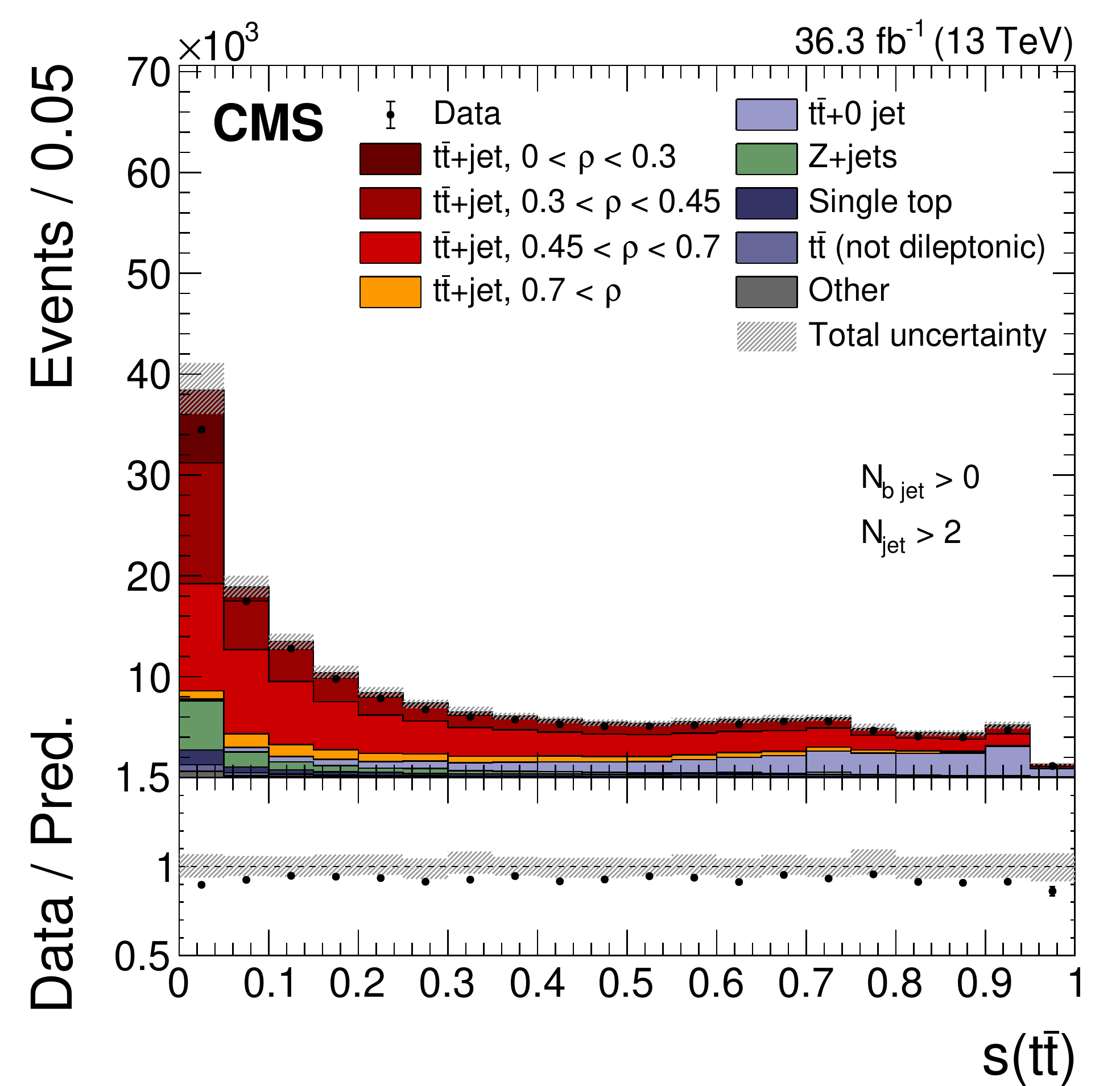}
    \caption{The observed (points) and MC predicted (stacked histograms) signal and background yields as a function of the signal (left) and \ttbarnojet background (right) output node score of the classifier NN. The vertical bars on the points represent the statistical uncertainty in the data. The hatched band represents the total uncertainty in the sum of the simulated signal and background predictions. The lower panels show the ratio of the data to the sum of the signal and background predictions.}
    \label{fig:DNNresponse}
\end{figure}

To maximize the signal sensitivity and increase the separation between the \ttbarjet signal and \Zjets and \ttbarnojet backgrounds, an MVA event classifier is developed. Using the same interface as for the regression NN, a fully connected feed-forward event-classification NN is defined, with the same strategies for the input-variable selection and optimization. Independent sets of training samples are obtained by using the NLO \MGvATNLO[{\FXFX}]+\PYTHIAEight simulation for the \ttbar and \Zjets events, which are not used in the likelihood fit. The input variables chosen for this NN, ranked by impact, are
\begin{itemize}
    \item  \pt of the additional third jet beyond the two jets from the top quark decays, as determined in the solution of the full kinematic reconstruction,
    \item  \pt of the third-highest-\pt jet,
    \item  invariant mass of the dilepton system,
    \item  \ptmiss,
    \item  mass of the leading lepton,
    \item  mass of the subleading lepton,
    \item  \pt of the leading jet,
    \item  \pt of the dilepton system,
    \item  number of reconstructed jets,
    \item  \pt of the leading lepton.
\end{itemize}
The architecture of the classification network has three output nodes, corresponding to the three process classes (\ttbarjet, \Zjets, and \ttbarnojet). The separation into multiple processes provides a better discrimination of the signal versus background events as compared to a binary-classifier approach. Because the response of the output nodes is used at a later step in the unfolding procedure (Section~\ref{sec:fit}), any bias in the obtained response with respect to the $\rho$ observable is reduced by making use of unsupervised domain adaptation by back propagation~\cite{bib:gradientReversal}.
In this method, additional layers are added to the NN with the optimization target of regressing \rhoreco as an ancillary output node, whereas the sign of the gradient in the back-propagation algorithm is inverted such that the NN does not learn to use input information sensitive to \rhoreco.
The trained NN uses four hidden layers with 204 nodes each.
The background rejection of the event-classification NN is shown in Fig.~\ref{fig:DNNresponse} by the distributions of the NN output scores for the \ttbarjet signal \sttj (left) and \ttbarnojet background $s(\ttbar)$ (right) events from data and simulation.

\section{Signal extraction, cross section measurement, and unfolding}
\label{sec:fit}
The differential cross section \dsigttjdrho is measured at the parton level in four $\rho$ bins using a maximum likelihood fit to final-state distributions, following the method of Ref.~\cite{bib:running}.
Since this method allows us to deal with larger background contributions, looser selection requirements compared to other analyses can be imposed. This allows us to maximize the signal acceptance and hence minimize the extrapolation uncertainties.
Systematic uncertainties are profiled in the likelihood function as nuisance parameters and are constrained using the data.
The expected signal and background distributions and the effect of the systematic uncertainties are estimated using the MC simulations.
The total cross section in each bin $k$ of the $\rho$ distribution, \sigttjk, is measured. This is related to the differential cross section \dsigttjdrho as
\begin{linenomath}\begin{equation}
    \sigttjk = \int_{\rhoklow}^{\rhokhigh} \dd{\sigttj}{\rho}\,\rd\rho,
\end{equation}\end{linenomath}
where \rhoklow and \rhokhigh denote the lower and the upper bounds of the $k$-th generator-level bin in $\rho$, respectively.

The likelihood function \likelihood assumes that the number of observed events in each bin follows a Poisson distribution and can be written as:
\begin{linenomath}\begin{equation}
    \likelihood = \prod_i \frac{\re^{-\vi} \vi^{\numi}}{\numi!} \prod_j \pi(\wj) \prod_m \pi(\lm).
\end{equation}\end{linenomath}
Here, $i$ denotes the index of the bin in the final-state distribution, and \vi and \numi are the expected and observed number of events in bin $i$, respectively.
The additional terms $\pi(\wj)$ and $\pi(\lm)$ are the prior probability density functions (pdfs) for the normalization of each background process \wj and nuisance parameter \lm, respectively.

The expected number of events \vi can be written as:
\begin{linenomath}\begin{equation}
    \vi = \sum_k \sik\left(\sigttjk,\,\vecl,\,\mtopMC\right) + \sum_j \bij\left(\wj,\,\vecl,\,\mtopMC\right),
\end{equation}\end{linenomath}
where \sik is the number of expected \ttbarjet signal events from the $k$-th generator-level bin for a reconstructed $\rho$ value in bin $i$. The value of \sik depends on \sigttjk, the nuisance parameters \vecl, and the top quark mass used in the simulation \mtopMC, which is needed to model the acceptance and extrapolate the results to the full phase space. Analogously, \bij is the number of expected background events from process $j$, and depends on its normalization \wj, the nuisance parameters \vecl, and on \mtopMC in the case of the \ttbar and \tW backgrounds.

Since \vi implicitly incorporates the dependence of the parton-level cross section \sigttjk on the experimental acceptance and detector response, the maximization of the likelihood function yields the results that are directly unfolded to the parton level.

For each cross section bin $k$, a signal strength parameter \rk is assigned, defined as
\begin{linenomath}\begin{equation}
    \rk = \frac{\sigttjk}{\sigttjkMC},
\end{equation}\end{linenomath}
where the denominator is the cross section corresponding to the normalization of the MC simulation.
In the fit, all the signal strength parameters are fitted simultaneously.
The \sigttjkMC values are determined using the nominal NLO \POWHEGPYTHIAEight simulation, and the nuisance parameters are constrained simultaneously with the differential cross section values.

To mitigate the correlation between the fitted signal strength parameters and \mtopMC, the latter is added as an additional free parameter to the fit.
For this purpose, additional predictions are obtained from two dedicated MC simulations where the value of \mtopMC is set to 169.5 or 175.5\GeV.
In the simultaneous fit, the dependence of the measured \sigttjk on \mtopMC is fully taken into account.
Therefore, the resulting \dsigttjdrho can be compared to the fixed-order calculations without assumptions on the relationship between \mtopMC and \mtpole ~\cite{bib:Kieseler:2015jzh}.
This method was proposed in Refs.~\cite{bib:Kieseler:2015jzh,bib:ttbarMTMCCorr} and was used in previous CMS measurements~\cite{bib:TillMatteo,bib:running}.
The likelihood construction and the minimization of $-2\ln(\likelihood)$ using \textsc{minuit}~\cite{bib:MINUIT} follows the procedures described in Refs.~\cite{bib:combineNote,bib:combineCMSHiggs,bib:combineCMSATLASHiggs}, as does the estimation of the uncertainties using \textsc{minos}~\cite{bib:MINUIT}.
The nuisance parameters corresponding to the shape and normalization are modeled using Gaussian and log-normal prior pdfs, respectively, based on their input uncertainties before the fit to the data.
The dependence of the templates on the nuisance parameters is interpolated quadratically (linearly) within (beyond) one standard deviation, as described in Ref.~\cite{bib:barlow2}.
The systematic uncertainty templates are built so as to reflect the variations found in estimating the systematic uncertainties.

The measurement is performed in four bins of \rhotrue and \rhoreco: 0--0.3, 0.3--0.45, 0.45--0.7, and 0.7--1.0, where the bin boundaries are determined by considering the experimental resolution and migration effects.
The bin edges are chosen so that the purity and stability of each bin is $\geq$50\%.
The bin edges for the $\rho$ variable correspond to about 1100, 750, 485, and 340\GeV for \mttbarjet.
The purity (stability) is defined as the number of signal events that are generated and reconstructed in the same bin, divided by the total number of signal events generated (reconstructed) in that bin.
The unfolding problem is determined to be well conditioned, and therefore no regularization is required.
To check for a possible bias of the fit, tests were performed where the true cross section was varied by a factor of two up and down for each parton-level bin, and the fit was repeated for each toy distribution. The fit was found to be unbiased.

\subsection{Event categorization}

The sensitivity of the cross section measurement to the different signal processes is enhanced by splitting the reconstructed events into multiple categories based on the jet ($\nj=1$, $\nj=2$, and $\nj\geq3$) and \PQb jet multiplicities ($\nbj=1$ and $\nbj\geq2$) and the \rhoreco bin. The different event types are labeled as ``mjnb'' for events with m jets and n \PQb jets.
Events with fewer than three reconstructed jets, corresponding mainly to background \ttbarnojet and \Zjets production, are also included in the fit.
This maximizes the acceptance of the events reconstructed in the visible phase space and helps to constrain the background.
Final-state distributions for each category are chosen that maximize the signal sensitivity.
The relative signal response \dnnresponse of the NN classifier, defined as: $\dnnresponse=\sttj / [\sttj+s(\ttbarnojet)]$, is used as the observable for event categories with more than two reconstructed jets.
To increase the fit sensitivity to \mtopMC, the minimum invariant mass \mlbmin among the lepton and \PQb jet combinations is used as the observable for events with two jets, where both are \PQb tagged.
In the remaining categories, the \pt of the lowest-\pt jet is fitted.
The three dileptonic channels are kept separate in the fit and are treated as independent categories to disentangle the effects of systematic uncertainties in the different lepton flavors.
An overview of the event categories and the chosen distributions are given in Table~\ref{tab:categories}.
Good agreement between the distributions from data and simulation is found.

\begin{figure}[!ht]
  \centering
  \includegraphics[width=.99\textwidth]{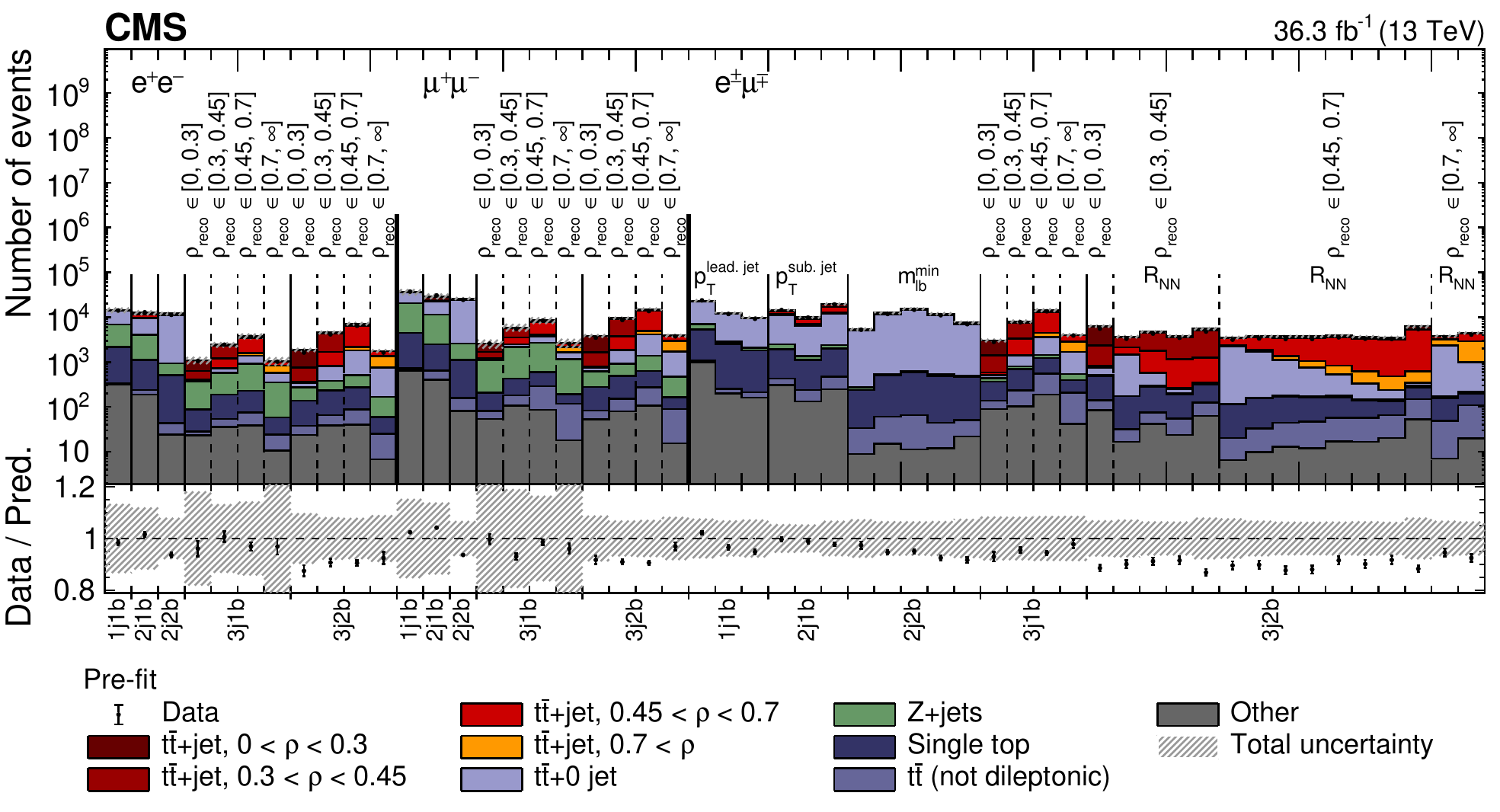}
  \caption{
  The distributions from data (points) and simulated signal and background (colored histograms) used in the maximum likelihood fits before the fit to the data. The distributions are shown for each dilepton type and each event category, where the x-axis label ``mjnb'' refers to events with m jets and n \PQb jets. The vertical bars on the points show the statistical uncertainty in the data. The hatched band represents the total uncertainty in the sum of the simulated signal and background predictions. The lower panel gives the ratio of the data to the sum of the simulated predictions.}
 \label{fig:prefit}
\end{figure}

\begin{table}[!ht]
    \centering \topcaption{A list of the event categories and distributions used in the maximum likelihood fit.} \label{tab:categories}\vspace{0.1cm}
\begin{tabular}{c|cccc|cc}
      & \multicolumn{4}{c|}{Reconstructed $\rho$} & \multicolumn{2}{c}{No reconstructed $\rho$}                           \\
      & \multicolumn{4}{c|}{$\nj\geq3$}  & \multirow{2}{*}{$\nj=1$} & \multirow{2}{*}{$\nj=2$} \\
      & $\rho<0.3$    & $0.3<\rho<0.45$    & $0.45<\rho<0.7$  & $\rho>0.7$ &                           &                         \\\hline
$\nbj=1$ & \dnnresponse     & \dnnresponse     & \dnnresponse   & \dnnresponse & $\pt^{\text{leading jet}}$                       & $\pt^{\text{subleading jet}}$                    \\
$\nbj\geq2$ & \dnnresponse     & \dnnresponse     & \dnnresponse  &  \dnnresponse & \NA                         & \mlbmin
\end{tabular}
\end{table}
In the binning of each observable, the statistical uncertainty in the nominal and systematic uncertainty templates is considered to reduce its impact on the extracted values of \rk, \wj, and \vecl after the fit to data.
In particular, the kinematic distributions listed in Table~\ref{tab:categories} are reduced to the total event yield if the number of simulated events is insufficient to model the kinematic distributions properly. As a criterion, the statistical uncertainty in the number of simulated events is not allowed to exceed 0.5\% per bin.
The final input distributions to the fit are displayed in Fig.~\ref{fig:prefit}, where the data are compared to the simulated signal and background distributions before the fit.

\section{Systematic uncertainties}
\label{sec:systematics}
Contributions to the systematic uncertainties from various sources are modeled as nuisance parameters in the fit, as described in Section~\ref{sec:fit}. For each variation, dedicated templates are obtained describing the effect on each background source or signal contribution.

\subsection{Experimental uncertainties}
Most of the experimental systematic uncertainties are estimated by varying the scale factors (SFs) used to correct any differences between the data and simulation. The following experimental uncertainties are considered:
\begin{itemize}
    \item The integrated luminosity used to normalize the simulated samples has a relative uncertainty of 1.2\%~\cite{bib:lumipaper16}.
    \item The uncertainty in the amount of pileup is estimated by varying the total inelastic \pp cross section by its measurement uncertainty of 4.6\%~\cite{bib:minbias} in the simulation.
    \item To correct for differences in the trigger efficiencies between data and simulation, dedicated SFs are derived in \pt and \sigeta bins of the leading and subleading leptons. These SFs are typically close to unity and are varied within their uncertainties, which are composed of statistical and systematic components are smaller than 3\%.
    \item Electron and muon identification and isolation efficiencies are measured in bins of the lepton \pt and \sigeta~\cite{bib:egamma,Sirunyan:2018}.
    Corresponding uncertainties in the SFs are typically on the order of 2--5\% for electrons, and 0.5--1.5\% for muons. They are varied within their uncertainties in the simulation to estimate the corresponding systematic uncertainty.
    \item Uncertainties due to correcting the electron and muon energy scales and resolutions are assessed separately by varying them within their uncertainties in the simulation.
    The typical uncertainty in the energy resolution is on the order of 0.5\% and 5\% for electrons and muons, respectively~\cite{bib:egamma,Sirunyan:2018}. The energy variations are propagated to \ptvecmiss.
    \item Several uncertainties affecting the correction of the jet energy scale (JES) are split into a set of 23 individual contributions~\cite{Khachatryan:2016kdb}. They are varied as a function of the jet \pt and \sigeta, and the effect is propagated to \ptvecmiss. The total uncertainty in the JES ranges from 1.0--3.5\% depending on the jet kinematic variables~\cite{Khachatryan:2016kdb}.
    \item An additional uncertainty in \ptvecmiss is derived by varying the energies of reconstructed particles not clustered into jets. Its effect on the resolution is on the order of 5--30\%~\cite{CMS-PAS-JME-16-004}.
    \item The effect from the jet energy resolution (JER) is estimated by varying the SFs for the jet four-momenta within their uncertainty on the order of 2--8\%, depending on the \sigeta region~\cite{Khachatryan:2016kdb}.
    \item Uncertainties in the correction of the simulation coming from the application of the pileup jet identification are estimated by varying the efficiency and mistagging rate within their uncertainties~\cite{bib:pileupjetid,bib:pileupJetIDRun2DPS}.
    They range between 5--30\%, depending on the jet \sigeta and \pt.
    \item During the 2016 data taking, a gradual shift in the timing of the inputs to the ECAL L1 trigger in the region $\abseta > 2.0$ caused a specific trigger inefficiency. For events containing an electron (jet) with \pt larger than $\approx$50 (100)\GeV, in the region $2.5 < \abseta < 3.0$ the efficiency loss is $\approx$10--20\%, depending on \pt, \sigeta, and the data-taking period~\cite{Sirunyan:2020zal}.
    A similar effect was present for L1-trigger muons because of the finite time resolution of the muon detectors, leading to an effect of 0.5--1.5\% per muon~\cite{CMS:2021yvr}. Dedicated SFs are derived for correcting the simulation for this effect known as ``prefiring'' and are varied within their uncertainties.
    \item To correct for different \PQb tagging efficiencies and mistagging rates of light-quark and gluon jets in the data and simulation, SFs are derived using simulated QCD multijet events and are applied as a function of the jet \pt~\cite{bib:btv}. To estimate the impact of that systematic uncertainty, the SFs are varied within their estimated uncertainties, which are split into individual subsources. For light- and heavy-quark jets the uncertainty ranges from 5-10 and 1-5\%, respectively.
\end{itemize}

\subsection{Theoretical uncertainties}
Additional systematic uncertainties arise from assumptions on the model parameters of the nominal \POWHEGPYTHIAEight simulation. Their effect is estimated by appropriate variations in the simulated samples. The following theoretical uncertainties are considered:
\begin{itemize}
    \item To estimate the effect of missing higher-order corrections in the NLO simulation, the \muR and \muF scales are varied by a factor of two up and down in the \POWHEG calculation with respect to their nominal values. Combined variations of \muR and \muF are not considered.
    \item Uncertainties in the modeling of the parton shower are assessed by varying individually the corresponding scales for the initial- (ISR) and final-state radiation (FSR) by a factor of two up and down with respect to their nominal values.
    \item The PDF uncertainty is estimated by using the 100 eigenvector variations of the NNPDF3.1 PDF set~\cite{bib:NNPDF1,bib:NNPDF2}, each treated as an individual nuisance parameter in the fit. Additionally, the value of $\alpS (\Zmass)$, where \Zmass is the \PZ boson mass, is varied within its uncertainty in this PDF set.
    \item The dependence on the matching between the ME and PS generators is estimated by varying the \hdamp parameter in the \POWHEG simulation within the tuning uncertainties~\cite{bib:cp5} in dedicated simulated samples.
    \item The modeling dependence on the CP5 UE tune is evaluated by varying the tune parameters within their uncertainties~\cite{bib:cp5}, for which dedicated samples are generated.
    \item In the \PYTHIAEight setup used for the nominal simulation, certain assumptions are made on the modeling of color reconnection (CR). While early-resonance decays (ERD) are turned off in the nominal setting, a dedicated sample with ERD enabled is generated, and the difference is treated as a systematic uncertainty. Moreover, alternative CR models are considered, including a gluon-move scheme~\cite{bib:gluonmove} and a QCD-inspired scheme~\cite{bib:qcdinspired}.
    \item Uncertainties originating from the limited knowledge of the \PQb hadron fragmentation function are estimated by varying the parameters of the Bowler--Lund function within their uncertainties~\cite{bib:bowler}. Alternatively, the Peterson fragmentation function~\cite{bib:peterson} is used.
    \item The limited precision on the semileptonic branching fractions of \PQb hadrons is estimated by varying them within their uncertainties, as estimated in Ref.~\cite{bib:precision5}.
    \item Differential measurements of the \ttbar cross section at $\sqrt{s}=13\TeV$ have demonstrated that the \pt distribution of the top quark is softer than predicted by the \POWHEG simulation~\cite{bib:kinreco,bib:ttbar1,bib:ttbar2,bib:ttbar3,bib:reweighting3}. An additional uncertainty, which is estimated by reweighting the simulation to the measured top quark \pt in Refs.~\cite{bib:reweighting1,bib:reweighting4}, is considered.

\end{itemize}
Besides the PDF and \hdamp variations, all the above-mentioned uncertainties are also assessed for \tW production and are considered correlated, if applicable. In the case of \Zjets production, \muR and \muF variations are also considered.
The theoretical uncertainties in the extrapolation to the full phase space are estimated by evaluating their impact on the signal acceptance following the strategy of Refs.~\cite{bib:running,bib:TillMatteo}.
The extrapolation uncertainty is determined after varying each nuisance parameter individually according to their uncertainties before the fit and by recomputing the cross section and evaluating the difference with respect to the nominal result.
The extrapolation uncertainty is determined as described in the following. Each relevant nuisance parameter $j$ is set to values corresponding to one standard deviation up and down while all others are fixed to the values after the fit. The resulting changes in the acceptance are recorded. The corresponding variations of the cross section values \sigttjk with respect to the nominal values are considered as the additional relative extrapolation uncertainty.

\subsection{Additional uncertainties}
Additional uncertainties arise from the normalizations of the background processes, the finite number of events in the MC simulation, and the corresponding effects on the templates from these sources.

The uncertainties in the normalizations for single top quark production and smaller background contributions, such as diboson and \Wjets production, are taken to be 30\% and modeled with a log-normal prior pdf, following the prescriptions from previous analyses~\cite{bib:ttbar3,bib:sasha_multiDim,bib:TillMatteo}.
For the \Zjets background, separate uncertainties are used for each \PQb jet category in order to remove the dependence of the fit result on the prediction of the \PQb jet multiplicity distribution from the \Zjets simulation. Similarly, the \Zjets background
is assigned an additional uncertainty of 5, 10, 30, and 50\% for events with exactly 0, 1, 2, and 3 or more jets, respectively. The first three uncertainties are estimated by performing scale variations in the \Wjets predictions with NLO precision, whereas the last one is assigned conservatively~\cite{bib:TillMatteo}.
For the \ttbarnojet background, no prior pdf is used since the normalization is left free in the fit in order to help constrain it with the background-dominated categories.

The statistical uncertainty due to the finite size of the MC event samples is evaluated via bin-by-bin nuisance parameters with the Barlow--Beeston ``light'' method~\cite{bib:barlow1,bib:barlow2}.
In addition, the impact of the finite statistical precision on the predictions of the MC samples with dedicated systematic variations is cross-checked using toy experiments, following the approach described in Ref.~\cite{bib:running}, and was found to be negligible.

\section{Results and extraction of the top quark mass}
\label{sec:mass}
\begin{figure}[!pth]
  \centering
  \includegraphics[width=.99\textwidth]{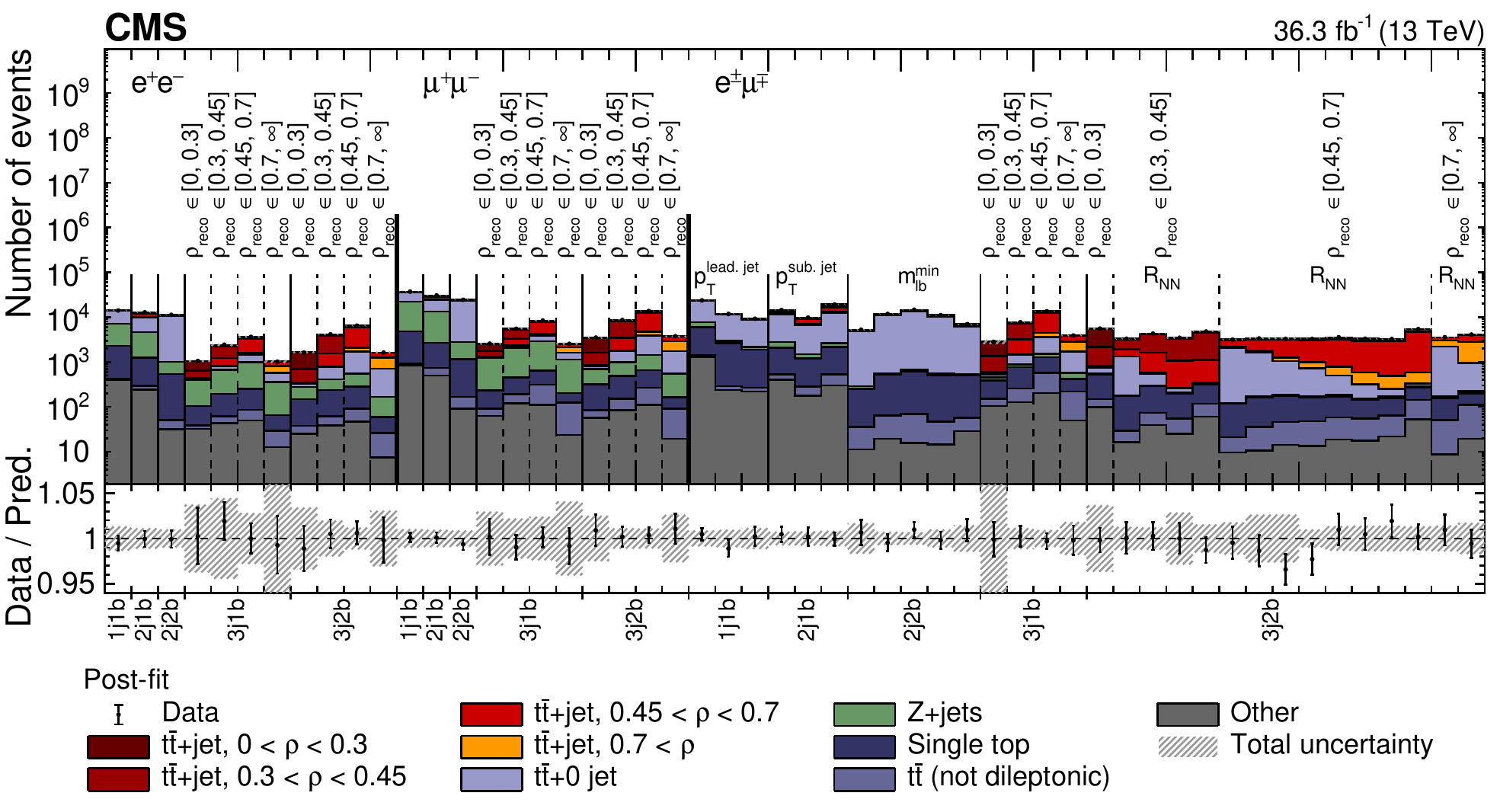}
  \caption{The distributions from data (points) and simulated signal and background (colored histograms) used in the maximum likelihood fits after the fit to the data. The distributions are shown for each dilepton type and each event category, where the x-axis label ``$m$j$n$b'' refers to events with $m$ jets and $n$ \PQb jets. The vertical bars on the points show the statistical uncertainty in the data. The hatched band represents the total uncertainty in the sum of the simulated signal and background predictions. The lower panel gives the ratio of the data to the sum of the simulated predictions.}
 \label{fig:postfit}
\end{figure}

\begin{figure}[!pth]
  \centering
  \includegraphics[width=.49\textwidth]{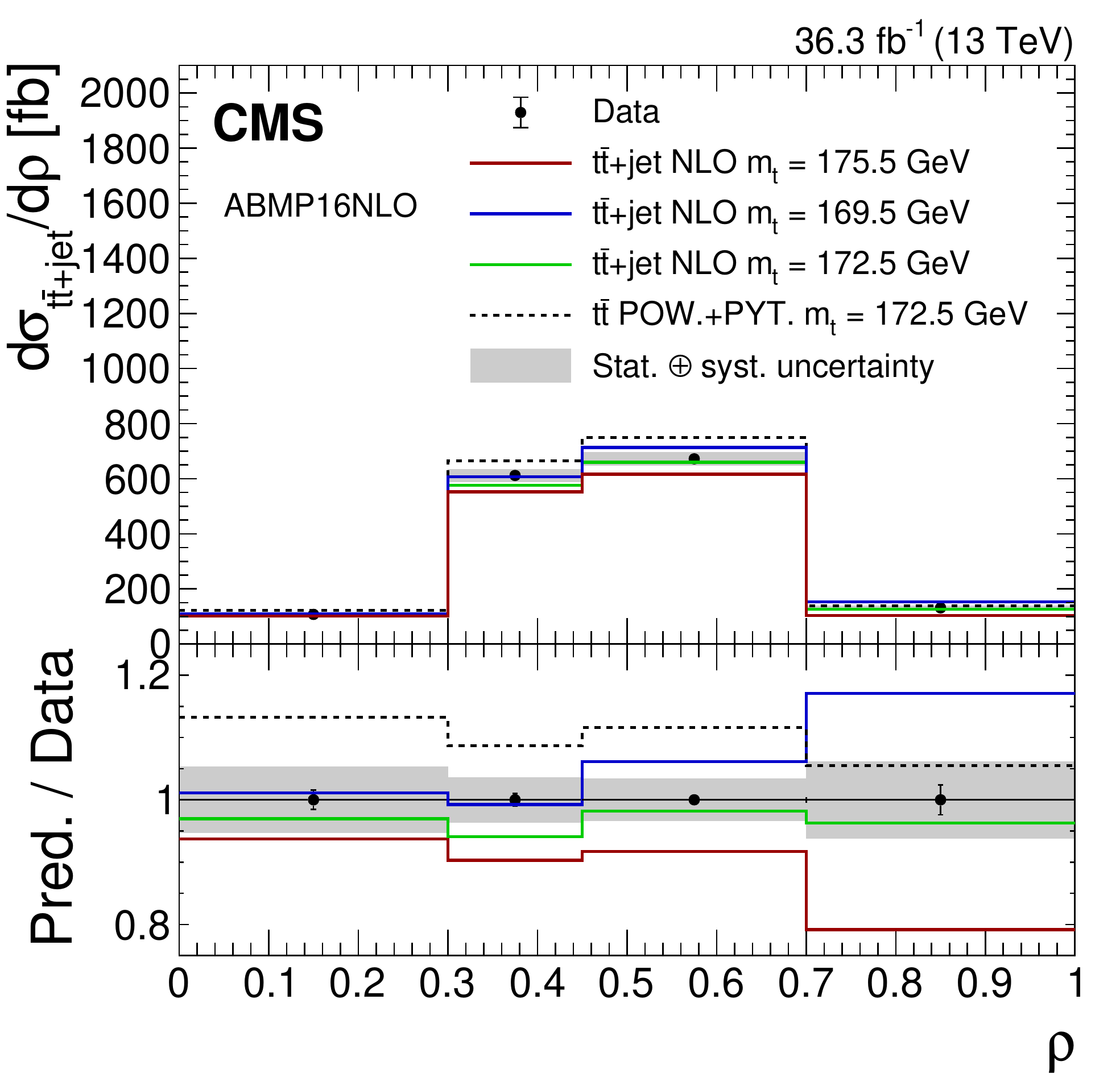}
  \hfill
  \includegraphics[width=.49\textwidth]{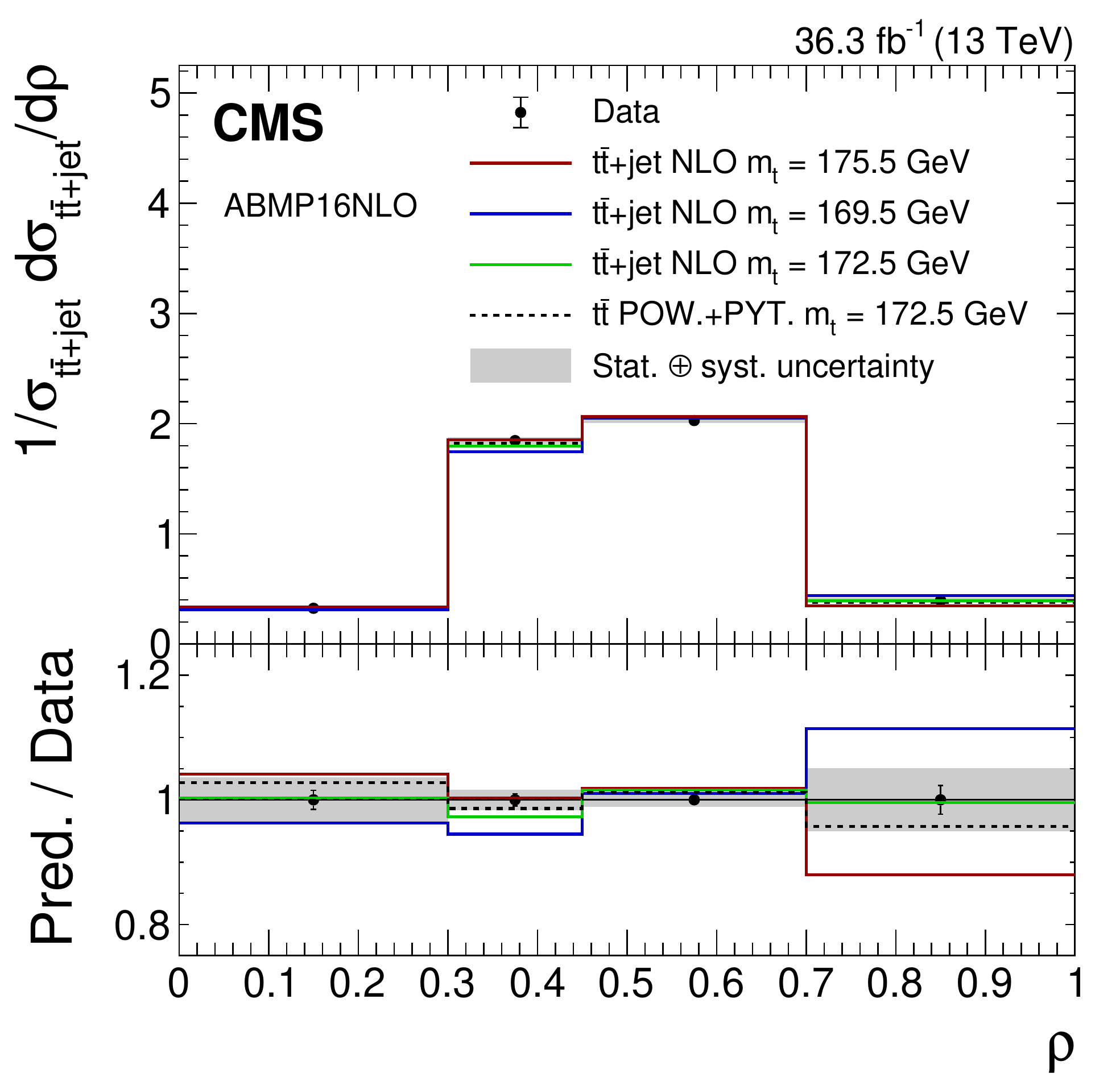}
  \caption{The absolute (left) and normalized (right) \ttbarjet differential cross section as a function of $\rho$ for the data (points) and theoretical predictions described in the text using the AMBP16NLO PDF set from the NLO MC with three different \mtop values and from the \POWHEG (POW) + \PYTHIAEight (PYT) calculations (lines). The vertical bars on the points show the statistical uncertainty in the data and the shaded region represents the total uncertainty in the measurement. The lower panels give the ratio of the predictions to the data.}
 \label{fig:diffXsec}
\end{figure}

Figure~\ref{fig:postfit} compares the resulting data distributions after the fit to the MC signal and background predictions.
The measured absolute differential cross section \dsigttjdrho is shown in Fig.~\ref{fig:diffXsec} (left) and compared to the NLO QCD predictions. The latter are obtained using the \ttbarjet process implemented in \POWHEG-\textsc{box}~\cite{bib:ttjPowheg}, with the ABMP16NLO~\cite{bib:ABMP16} PDF set, and assuming \mtpole values of 169.5, 172.5, and 175.5\GeV.
The same kinematic requirements on the additional jet as defined in Section~\ref{subsec:partonlevel} are used, while the simulation is not interfaced to any parton shower.
It is worth mentioning that in the ABMP analysis, the PDFs and $\alpS (\Zmass)$ are obtained simultaneously at the same perturbative order in QCD.
Additionally, the CT18NLO PDF set~\cite{bib:CT18} is considered.

{\tolerance=800
The QCD scale in the theoretical prediction is dynamically set to $\HTB/2$, as suggested in Ref.~\cite{bib:DynScaleKrausWorek} and discussed in Ref.~\cite{bib:ttjPheno}, where \HTB is defined as the scalar sum of the top quark and antiquark transverse masses and the \pt of the additional jet.
The values and the uncertainty in the normalized differential cross section are derived from the fit results for the absolute cross section by taking into account the covariance matrix as obtained in the fit to data.
The normalized differential cross section is determined by dividing the value of the absolute differential cross section for each bin by the sum of the values of all the bins.
Uncertainties are symmetrized prior to the normalization as provided by \textsc{hesse}~\cite{bib:MINUIT}.
The results are shown in Fig.~\ref{fig:diffXsec} (right) in comparison with the theoretical predictions.
While the absolute cross section shows a stronger dependence on \alpS, this dependence is small for the normalized cross section~\cite{bib:ttjPheno}.
All the sources and values of the systematic uncertainties in the differential cross section result are given in Table~\ref{tab:uncertainties}.
\par}

\begin{table}[!p]
    \centering \topcaption{The relative uncertainties $\Delta{\sigttjk}$ in the parton-level cross section values \sigttjk and their sources in each bin $k$ of the $\rho$ distribution. The statistical uncertainty is evaluated by keeping all nuisance parameters fixed to their values after the fit to data. The breakdown of the uncertainty is obtained by repeating the fit after fixing the nuisance parameters related to the components under consideration to their fitted values. The partial uncertainty is then estimated by subtracting the uncertainty obtained with this procedure from the total uncertainty. The quadratic sum of the contributions is different from the total uncertainty because of correlations between the nuisance parameters.}
    \label{tab:uncertainties}
    \begin{tabular}{p{1pt}l@{}cccc}
    \multicolumn{2}{l}{Uncertainty Source} & $\Delta{\sigttj^{1}}$ [\%] & $\Delta{\sigttj^{2}}$ [\%] & $\Delta{\sigttj^{3}}$ [\%] & $\Delta{\sigttj^{4}}$ [\%] \\\hline
    \multicolumn{2}{l}{Experimental} &  &  &  & \\
    & Muon identification & 1.8 & 1.5 & 1.5 & 1.4\\
    & Muon energy scale and resolution & 0.7 & 0.2 & 0.3 & 0.5\\
    & Electron identification & 2.0 & 1.7 & 1.7 & 2.1\\
    & Electron energy scale and resolution& 0.9 & 1.0 & 0.9 & 1.5\\
    & Jet energy scale & 2.6 & 2.0 & 2.2 & 3.6\\
    & Jet energy resolution & 0.6 & 0.5 & 0.5 & 0.4\\
    & Jet identification & 1.1 & 0.8 & 0.8 & 1.3\\
    & \ptmiss & 0.2 & 0.3 & 0.4 & 0.8\\
    & \PQb jet identification & 1.0 & 0.7 & 0.6 & 1.2\\
    & Trigger efficiency & 1.8 & 1.2 & 1.1 & 1.8\\
    \multicolumn{2}{l}{Total} & 4.0 & 3.1 & 3.1 & 4.7\\[\cmsTabSkip]
    \multicolumn{2}{l}{Background normalization} &  &  &  & \\
    & \ttbarnojet & 2.2 & 2.0 & 1.7 & 0.7\\
    & \Zjets & 2.4 & 1.9 & 1.7 & 2.6\\
    & Single top quark & 0.9 & 0.8 & 0.7 & 0.1\\
    \multicolumn{2}{l}{Total} & 3.1 & 2.5 & 2.4 & 2.7\\[\cmsTabSkip]
    \multicolumn{2}{l}{Modeling} &  &  &  & \\
    & \Zjets ME scale & 0.7 & 0.4 & 0.2 & 0.3\\
    & Single top quark ME/FSR/ISR scales & 1.2 & 0.6 & 0.4 & 0.1\\
    & \ttbar PDF & 0.1 & 0.1 & 0.1 & 0.6\\
    & \ttbar ME scale & 1.0 & 0.5 & 0.6 & 0.4\\
    & \ttbar ISR scale & 1.2 & 0.8 & 0.6 & 1.6\\
    & \ttbar FSR scale & 1.3 & 0.8 & 0.6 & 1.7\\
    & \ttbar top quark \pt & 2.0 & 1.3 & 0.1 & 1.2\\
    & \PQb fragmentation & 0.9 & 0.7 & 0.8 & 0.8\\
    & Color reconnection & 0.5 & 0.6 & 0.2 & 0.7\\
    & \ttbar matching scale & 0.6 & 0.5 & 0.6 & $\leq$0.1\\
    & Underlying-event tune & 0.2 & 0.5 & 0.2 & 0.5\\
    \multicolumn{2}{l}{Total} & 3.2 & 1.9 & 1.8 & 3.1\\[\cmsTabSkip]
    \multicolumn{2}{l}{Integrated luminosity} & 1.2 & 1.4 & 1.3 & 1.2\\
    \multicolumn{2}{l}{\mtopMC} & 1.7 & 1.0 & 0.4 & 2.3\\
    \multicolumn{2}{l}{Finite size of simulated samples} & 2.0 & 1.4 & 1.3 & 2.2\\[\cmsTabSkip]
    \multicolumn{2}{l}{Total systematic} & 5.0 & 3.4 & 3.2 & 5.6\\
    \multicolumn{2}{l}{Statistical} & 1.6 & 1.0 & 0.8 & 2.4\\[\cmsTabSkip]
    \multicolumn{2}{l}{Total} & 5.2 & 3.6 & 3.3 & 6.1\\
    \end{tabular}
\end{table}

\begin{figure}[!ht]
  \centering
  \includegraphics[width=.99\textwidth]{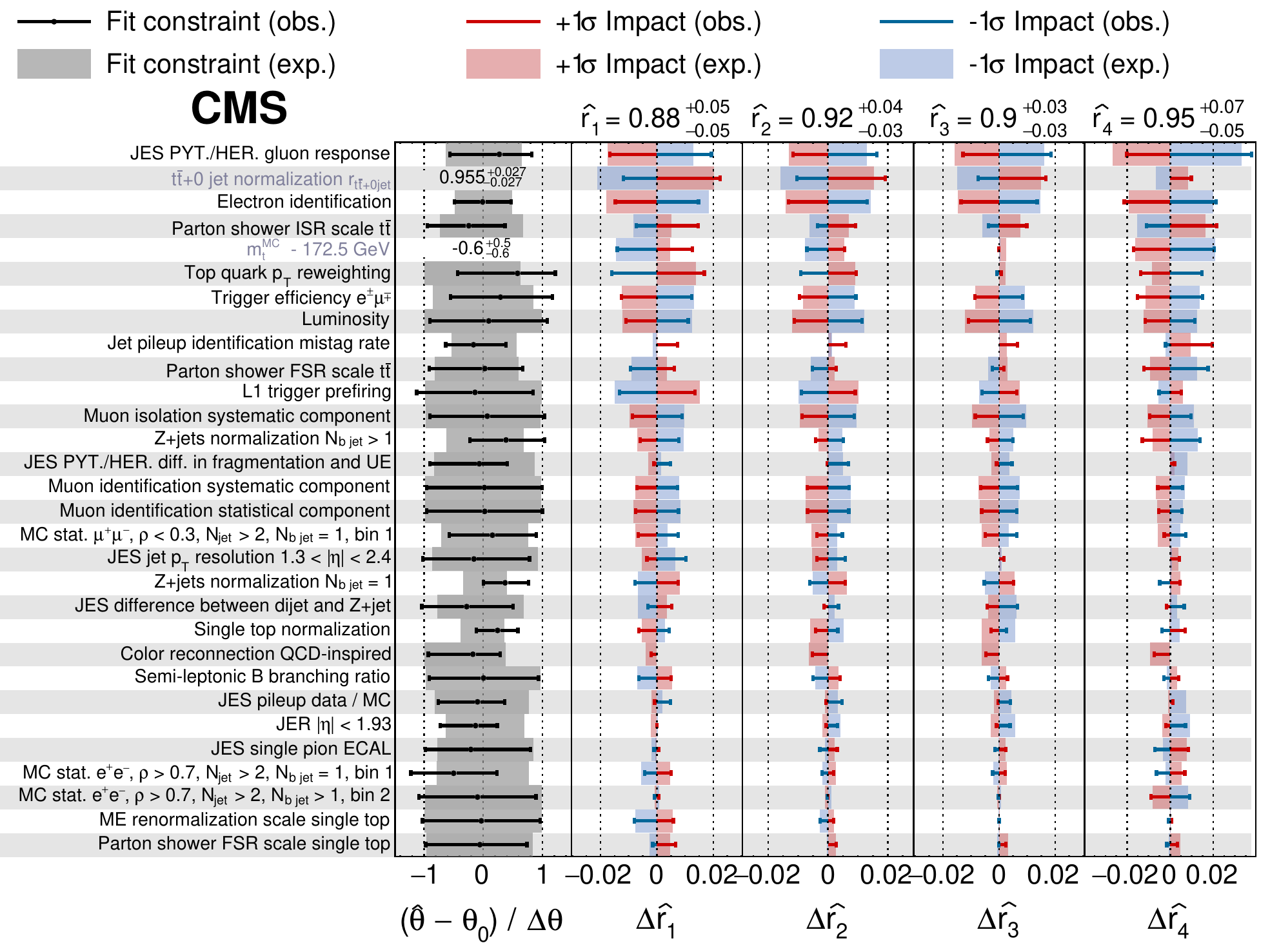}
  \caption{The fitted nuisance-parameter values and their impacts \dhatrk on the signal strengths \hatrk from the fit to the data, ordered by their relative summed impact. Only the 30 highest ranked parameters are shown. The resulting fitted values of \hatrk and their total uncertainties are also given. The nuisance-parameter values (\hattheta, black lines) are shown in comparison to their input values \thetaz before the fit and relative to their uncertainty \dtheta. The impact \dhatrk for each nuisance parameter is the difference between the nominal best fit value of \rk and the best fit value when only that nuisance parameter is set to its best fit value \hattheta while all others are left free. The red and blue lines correspond to the variation in \dhatrk when the nuisance parameter is varied up and down by its fitted uncertainty (\dtheta), respectively. The corresponding gray, red, and blue regions show the expected values from fits to pseudo-data. For the nuisance parameters associated with the \ttbarnojet normalization and \mtopMC, the values after the fit to the data are given, because no prior pdf is assigned.}
 \label{fig:impact}
\end{figure}

The fitted parameter values, expected and observed constraints, and impacts on the signal strengths \rk for the 30 most important nuisance parameters after the fit to data are shown in Fig.~\ref{fig:impact}.
For each parameter, the difference between the best fit (\hattheta) and input value (\thetaz) is given relative to the input uncertainty \dtheta, while the constraint is defined as the uncertainty after the fit relative to the input uncertainty.
For the nuisance parameters associated with the \ttbarnojet normalization and \mtopMC, the values of the nuisance parameters after the fit are given in the figure instead of plotting $(\hattheta - \thetaz)/\dtheta$ because no prior pdf is assigned.
The values of the nuisance parameters after the fit are compatible with the corresponding values before the fit within their prior uncertainties, reflecting the good agreement between data and simulation before the fit.
Some differences between \hattheta and \thetaz for nuisance parameters describing the background normalization for the single top quark and \Zjets processes are larger.
This is expected because of the limitations in the MC simulations used to model these backgrounds in the signal phase space region, which often corresponds to the tails of the background processes.
The nuisance parameter associated with the modeling of the top quark \pt spectra also has a larger value after the fit. This is because of the different spectra observed in data and simulation, as discussed in Section~\ref{sec:eventSelection}.

The observed constraints and impacts agree with the expectations derived from fitting pseudo-data obtained from the nominal simulation.
Strong constraints are determined on the normalizations of the background \Zjets and single top quark processes, which are explained by their large number of events in the background-dominated categories that helps to constrain the large normalization uncertainty.
The nuisance parameter associated with the electron identification efficiency has a tight constraint from combining decay channels with different electron and muon multiplicities in the fit. This effect was seen in previous CMS measurements~\cite{bib:TillMatteo,bib:running}.

The constraints on nuisance parameters associated with the jet energy corrections, JER, and jet identification demonstrate the power of the fit method, since the analysis phase space region and fitted kinematic distributions are sensitive to these effects. Also, because the analyzed phase space region is different from the one used to derive the corresponding calibrations described in Section~\ref{sec:systematics}, deviations from the values before the fit and constraints are expected.
In the fit, \mtopMC is determined as $171.93 \pm 0.65\GeV$, where the uncertainty includes both the statistical and systematic components.

\begin{figure}[!bh]
  \centering
  \includegraphics[width=.49\textwidth]{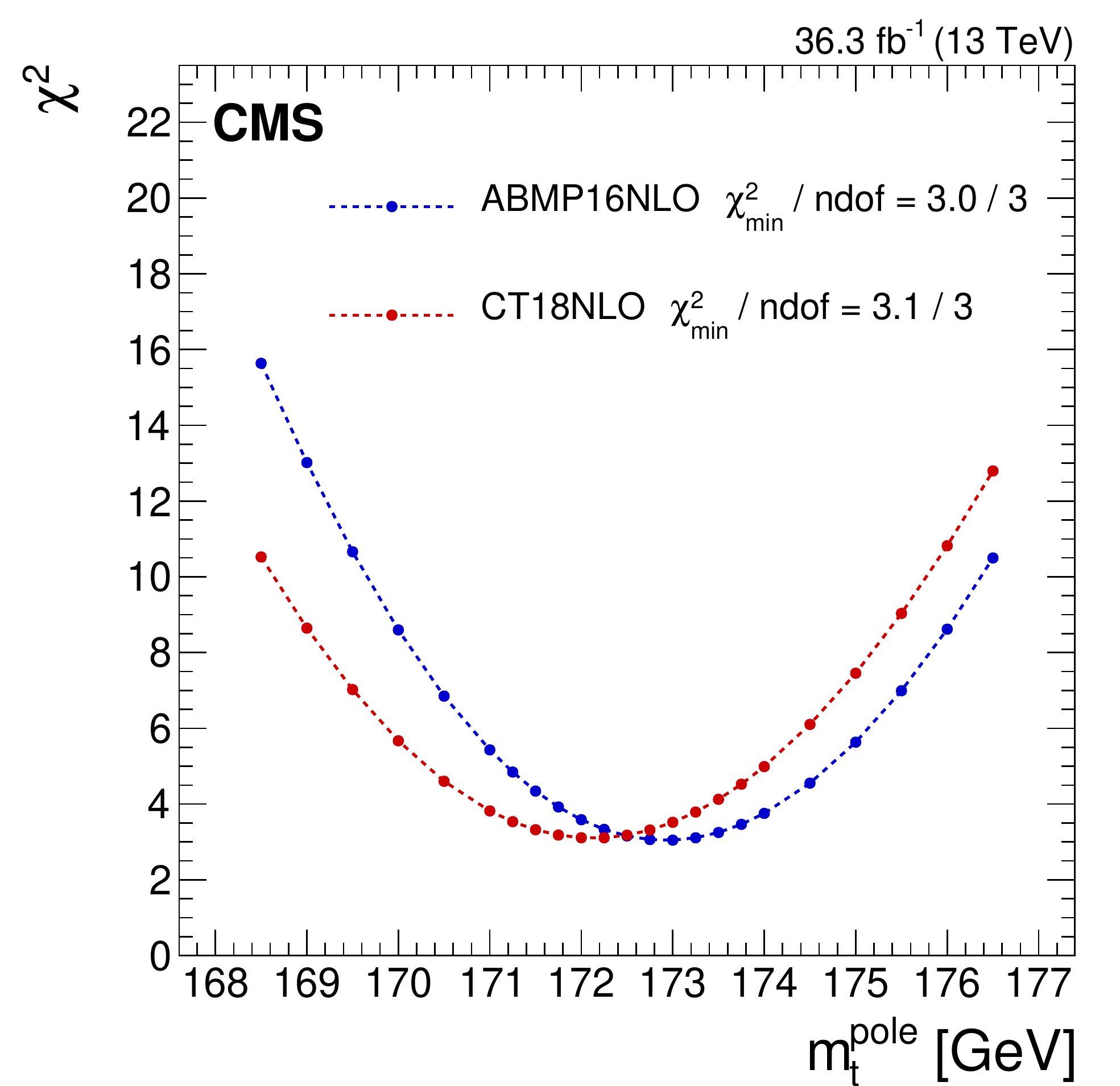}
  \hfill
  \includegraphics[width=.49\textwidth]{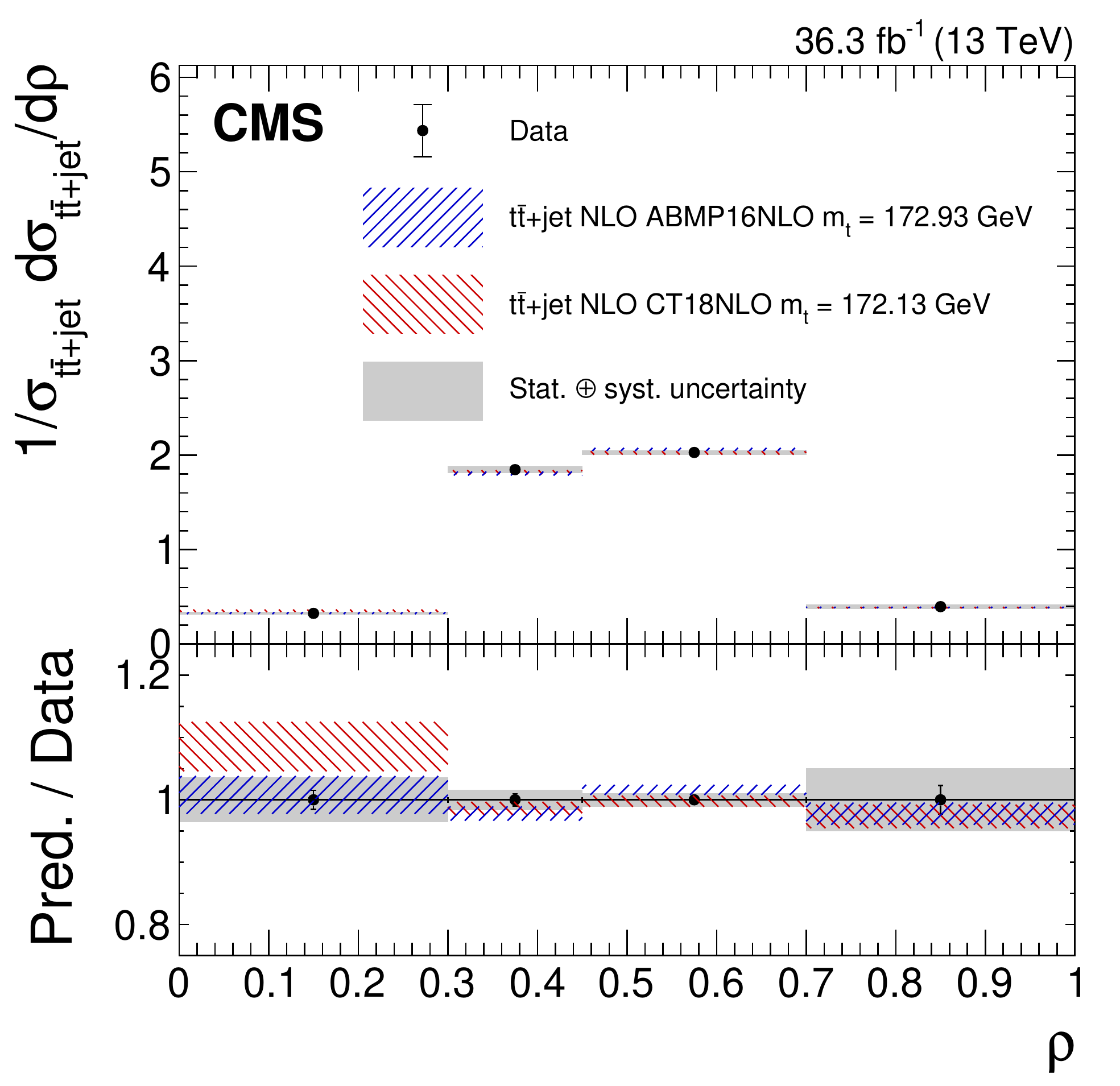}
  \caption{Left: The \chisq values versus \mtpole from the fit of the measured normalized \ttbarjet differential cross sections to the theoretical predictions using the ABMP16NLO (blue points) and CT18NLO (red points) PDF sets. The minimum \chisq value and the number of degrees of freedom (ndof) are given for each fit. Right: The measured normalized \ttbarjet differential cross section (points) as a function of $\rho$, compared to the predictions using the two PDF sets and the corresponding best fit values for \mtpole (hatched bands). The lower panel gives the ratio of the theoretical predictions to the measured values. For both panels, the vertical bars on the points show the statistical uncertainty in the data, the height of the hatched bands represent the theoretical uncertainties in the predictions, and the gray band gives the total uncertainty in the measured cross section.}
 \label{fig:massfit}
\end{figure}

The value of \mtpole is extracted from a \chisq fit of the theoretical predictions for the normalized differential cross section at NLO to the measurement.
Since in a fit to a normalized distribution, the value of any bin is fully determined by the value of the others, one bin is removed from the \chisq fit.
The result is independent of the choice of the removed bin.
The covariance matrix used in the \chisq fit is derived by combining the one obtained in the likelihood fit, the extrapolation uncertainties in the measured cross section, and the PDF uncertainties in the theoretical predictions.
The PDF uncertainties are evaluated in each bin and included in the total covariance matrix.
For CT18NLO, the uncertainties evaluated at 90\% confidence level (\CL), are rescaled to the 68\% \CL for consistency with the ABMP16NLO PDF set and symmetrized by taking the maximum of the up and down variation.
The extrapolation uncertainties for all relevant theoretical uncertainties, as described in Section~\ref{sec:systematics}, are symmetrized by taking the average of the positive and negative impacts and are included in the calculation of the \chisq.
The resulting \chisq curves are shown in Fig.~\ref{fig:massfit} (left). The best fit value for \mtpole is extracted at the \chisq minimum and the experimental uncertainty is obtained using the tolerance criterion of $\Delta \chisq =1$.

To estimate the scale variation uncertainty, the fit is repeated using the theoretical predictions, with \muR and \muF varied independently by factors of 0.5 and 2, avoiding cases where $\muF/\muR = 4$ or $1/4$.
The total scale uncertainty is estimated by taking the maximum difference in the results to the nominal one.
The resulting value of
\begin{linenomath}\begin{equation*}
    \mtpole = 172.93\pm1.26\fit\,^{+0.51}_{-0.43}\scale\GeV
\end{equation*}\end{linenomath}
is obtained using the ABMP16NLO PDF set.
The first uncertainty corresponds to the total statistical and systematic uncertainties from the fit including the PDF and extrapolation uncertainties, and the second uncertainty comes from the variation in the \muR and \muF scales.
Using the CT18NLO PDF set instead, the measured value is
\begin{linenomath}\begin{equation*}
    \mtpole = 172.13\pm1.34\fit\,^{+0.50}_{-0.40}\scale\GeV.
\end{equation*}\end{linenomath}
The total uncertainty in \mtpole corresponds to 1.36 (1.43)\GeV for the ABMP16NLO (CT18NLO) PDF set. Comparisons between the unfolded data and theoretical predictions for the determined values of \mtpole with both PDF sets are given in Fig.~\ref{fig:massfit} (right), showing good agreement between the fitted prediction and the measured cross section for both PDF sets.
The impact of the individual PDF uncertainties is estimated by replacing the central values of the measured cross section with the ones obtained from the theoretical prediction, assuming a top quark mass of 172.5\GeV. Two fits are performed: one using the full covariance matrix, and the other with a covariance matrix that excludes the effect of the PDF uncertainties. The difference in quadrature of the two results represents an approximate estimate of the impact of the PDF uncertainties. This procedure yields a total impact of 0.35 (0.27) \GeV for the CT18NLO (ABMP16NLO) PDF set.

The obtained results are in good agreement with previous measurements of \mtpole using \ttbarjet events~\cite{bib:ttjetAtlas8} at a center-of-mass energy of 8\TeV by ATLAS and the triple-differential cross sections for \ttbar production at a center-of-mass energy of 13\TeV by CMS~\cite{bib:sasha_multiDim}.
Compared to the ATLAS result, a lower sensitivity to \mtpole is expected in this analysis due to the higher center-of-mass energy.
Furthermore, in this analysis the systematic uncertainty due to the PDFs is fully taken into account, which leads to an increase in the total uncertainty in \mtpole.

\section{Summary}
\label{sec:summary}

Measurements are presented of the normalized differential cross section of top quark-antiquark pair (\ttbar) production in association with at least one additional jet as a function of the inverse of the invariant mass of the \ttbarjet system $\rho=2\mzero/\mttbarjet$, with the scaling constant $\mzero=170\GeV$.
Proton-proton collision data collected by the CMS experiment at the CERN LHC at a center-of-mass energy of 13\TeV are used, corresponding to an integrated luminosity of 36.3\fbinv.
Events in the dilepton decay channel are considered, and a novel multivariate analysis technique is applied to maximize the sensitivity to the signal process.
The differential cross section is measured at the parton level using a maximum likelihood fit to final-state observables, where all systematic uncertainties are profiled.
The value of the top quark pole mass \mtpole is extracted by comparing the measured \ttbarjet normalized differential cross section as a function of $\rho$ to theoretical predictions at next-to-leading order in quantum chromodynamics, obtained with two sets of parton distribution functions.
The \mtpole values is determined to be $172.93\pm1.36\GeV$ and $172.13\pm1.43\GeV$ using the ABMP16NLO and CT18NLO parton distribution functions, respectively.
Here, the uncertainties shown include the total statistical and systematic uncertainties including extrapolation uncertainties, and the theoretical uncertainties from the parton distribution functions and the matrix-element scales.
The results are in good agreement with previous measurements.

\begin{acknowledgments}
We congratulate our colleagues in the CERN accelerator departments for the excellent performance of the LHC and thank the technical and administrative staffs at CERN and at other CMS institutes for their contributions to the success of the CMS effort. In addition, we gratefully acknowledge the computing centers and personnel of the Worldwide LHC Computing Grid and other centers for delivering so effectively the computing infrastructure essential to our analyses. Finally, we acknowledge the enduring support for the construction and operation of the LHC, the CMS detector, and the supporting computing infrastructure provided by the following funding agencies: BMBWF and FWF (Austria); FNRS and FWO (Belgium); CNPq, CAPES, FAPERJ, FAPERGS, and FAPESP (Brazil); MES and BNSF (Bulgaria); CERN; CAS, MoST, and NSFC (China); MINCIENCIAS (Colombia); MSES and CSF (Croatia); RIF (Cyprus); SENESCYT (Ecuador); MoER, ERC PUT and ERDF (Estonia); Academy of Finland, MEC, and HIP (Finland); CEA and CNRS/IN2P3 (France); BMBF, DFG, and HGF (Germany); GSRI (Greece); NKFIH (Hungary); DAE and DST (India); IPM (Iran); SFI (Ireland); INFN (Italy); MSIP and NRF (Republic of Korea); MES (Latvia); LAS (Lithuania); MOE and UM (Malaysia); BUAP, CINVESTAV, CONACYT, LNS, SEP, and UASLP-FAI (Mexico); MOS (Montenegro); MBIE (New Zealand); PAEC (Pakistan); MES and NSC (Poland); FCT (Portugal); MESTD (Serbia); MCIN/AEI and PCTI (Spain); MOSTR (Sri Lanka); Swiss Funding Agencies (Switzerland); MST (Taipei); MHESI and NSTDA (Thailand); TUBITAK and TENMAK (Turkey); NASU (Ukraine); STFC (United Kingdom); DOE and NSF (USA).
      
\hyphenation{Rachada-pisek} Individuals have received support from the Marie-Curie programme and the European Research Council and Horizon 2020 Grant, contract Nos.\ 675440, 724704, 752730, 758316, 765710, 824093, 884104, and COST Action CA16108 (European Union); the Leventis Foundation; the Alfred P.\ Sloan Foundation; the Alexander von Humboldt Foundation; the Belgian Federal Science Policy Office; the Fonds pour la Formation \`a la Recherche dans l'Industrie et dans l'Agriculture (FRIA-Belgium); the Agentschap voor Innovatie door Wetenschap en Technologie (IWT-Belgium); the F.R.S.-FNRS and FWO (Belgium) under the ``Excellence of Science -- EOS" -- be.h project n.\ 30820817; the Beijing Municipal Science \& Technology Commission, No. Z191100007219010; the Ministry of Education, Youth and Sports (MEYS) of the Czech Republic; the Hellenic Foundation for Research and Innovation (HFRI), Project Number 2288 (Greece); the Deutsche Forschungsgemeinschaft (DFG), under Germany's Excellence Strategy -- EXC 2121 ``Quantum Universe" -- 390833306, and under project number 400140256 - GRK2497; the Hungarian Academy of Sciences, the New National Excellence Program - \'UNKP, the NKFIH research grants K 124845, K 124850, K 128713, K 128786, K 129058, K 131991, K 133046, K 138136, K 143460, K 143477, 2020-2.2.1-ED-2021-00181, and TKP2021-NKTA-64 (Hungary); the Council of Science and Industrial Research, India; the Latvian Council of Science; the Ministry of Education and Science, project no. 2022/WK/14, and the National Science Center, contracts Opus 2021/41/B/ST2/01369 and 2021/43/B/ST2/01552 (Poland); the Funda\c{c}\~ao para a Ci\^encia e a Tecnologia, grant CEECIND/01334/2018 (Portugal); the National Priorities Research Program by Qatar National Research Fund; MCIN/AEI/10.13039/501100011033, ERDF ``a way of making Europe", and the Programa Estatal de Fomento de la Investigaci{\'o}n Cient{\'i}fica y T{\'e}cnica de Excelencia Mar\'{\i}a de Maeztu, grant MDM-2017-0765 and Programa Severo Ochoa del Principado de Asturias (Spain); the Chulalongkorn Academic into Its 2nd Century Project Advancement Project, and the National Science, Research and Innovation Fund via the Program Management Unit for Human Resources \& Institutional Development, Research and Innovation, grant B05F650021 (Thailand); the Kavli Foundation; the Nvidia Corporation; the SuperMicro Corporation; the Welch Foundation, contract C-1845; and the Weston Havens Foundation (USA).
\end{acknowledgments}

\bibliography{auto_generated}
\cleardoublepage \appendix\section{The CMS Collaboration \label{app:collab}}\begin{sloppypar}\hyphenpenalty=5000\widowpenalty=500\clubpenalty=5000
\cmsinstitute{Yerevan Physics Institute, Yerevan, Armenia}
{\tolerance=6000
A.~Tumasyan\cmsAuthorMark{1}\cmsorcid{0009-0000-0684-6742}
\par}
\cmsinstitute{Institut f\"{u}r Hochenergiephysik, Vienna, Austria}
{\tolerance=6000
W.~Adam\cmsorcid{0000-0001-9099-4341}, J.W.~Andrejkovic, T.~Bergauer\cmsorcid{0000-0002-5786-0293}, S.~Chatterjee\cmsorcid{0000-0003-2660-0349}, K.~Damanakis\cmsorcid{0000-0001-5389-2872}, M.~Dragicevic\cmsorcid{0000-0003-1967-6783}, A.~Escalante~Del~Valle\cmsorcid{0000-0002-9702-6359}, P.S.~Hussain\cmsorcid{0000-0002-4825-5278}, M.~Jeitler\cmsAuthorMark{2}\cmsorcid{0000-0002-5141-9560}, N.~Krammer\cmsorcid{0000-0002-0548-0985}, L.~Lechner\cmsorcid{0000-0002-3065-1141}, D.~Liko\cmsorcid{0000-0002-3380-473X}, I.~Mikulec\cmsorcid{0000-0003-0385-2746}, P.~Paulitsch, F.M.~Pitters, J.~Schieck\cmsAuthorMark{2}\cmsorcid{0000-0002-1058-8093}, R.~Sch\"{o}fbeck\cmsorcid{0000-0002-2332-8784}, D.~Schwarz\cmsorcid{0000-0002-3821-7331}, S.~Templ\cmsorcid{0000-0003-3137-5692}, W.~Waltenberger\cmsorcid{0000-0002-6215-7228}, C.-E.~Wulz\cmsAuthorMark{2}\cmsorcid{0000-0001-9226-5812}
% ...
\par}
\vskip\cmsinstskip
\dag:~Deceased\\
$^{1}$Also at Yerevan State University, Yerevan, Armenia\\
$^{2}$Also at TU Wien, Vienna, Austria\\
% ...
\end{sloppypar}
\end{document}